\documentclass[12pt,aps,amsmath,latexsym,amsfonts]{JHEP3}

\usepackage{epsfig}
\usepackage{amsfonts,amssymb,amsmath}
\usepackage[hang]{subfigure}

\newcommand{\be}{\begin{equation}}
\newcommand{\ee}{\end{equation}}
\newcommand{\bea}{\begin{align}}
\newcommand{\eea}{\end{align}}

\title{The Rich Structure of Gauss-Bonnet Holographic Superconductors}
\author{
Luke Barclay\thanks{Email: luke.barclay@durham.ac.uk} 

\\
{\it Centre for Particle Theory, Department of Mathematical Sciences\\
Durham University, South Road, Durham, DH1 3LE, UK}}

\abstract{ We study fully  backreacting, Gauss-Bonnet (GB) holographic superconductors in 5 bulk spacetime dimensions.  We explore the system's dependence on the scalar mass for both positive and negative GB coupling, $\alpha$.  We find that when the mass approaches the Breitenlohner-Freedman (BF) bound and $\alpha\rightarrow L^2/4$ the effect of backreaction is to increase the critical temperature, $T_c$, of the system: the opposite of its effect in the rest of parameter space.  We also find that reducing $\alpha$ below zero increases $T_c$ and that the effect of backreaction is diminished.  We study the zero temperature limit, proving that this system does not permit regular solutions for a non-trivial, tachyonic scalar field and constrain possible solutions for fields with positive masses.  We investigate singular, zero temperature solutions in the Einstein limit but find them to be incompatible with the concept of GB gravity being a perturbative expansion of Einstein gravity.  We study the conductivity of the system, finding that the inclusion of backreaction hinders the development of poles in the conductivity that are associated with quasi-normal modes approaching the real axis from elsewhere in the complex plane.
}

\preprint{DCPT-10/73}

\begin{document}
\newcommand{\zed}{$\mathbb{Z}_2$}

\section{Introduction}

The gauge/gravity correspondence \cite{Maldacena:1997re} provides a means of studying strongly coupled field theories via the analysis of more amenable, weakly coupled gravity theories.  In recent years the correspondence has been applied to condensed matter systems leading to interesting descriptions of apparent superconductors \cite{Herzog:2009xv,Hartnoll:2009sz,Horowitz:2010gk}.  Its application to superconductivity is of particular interest due to the discovery of, so called, high temperature superconductors in the 1980's which fall outside of the scope of current theories of superconductivity and are thought to be described by strongly coupled field theories. 

In their simplest manifestations, the gravitational duals to these superconducting systems consist of a black hole in an anti-de Sitter (adS) spacetime with a complex scalar field coupled to a U(1) gauge field.  An interesting characteristic of such theories is that they can remain stable despite the scalar field having a negative mass squared, provided it is not below the Breitenlohner-Freedman (BF) bound \cite{Breitenlohner:1982jf}.  However, it is argued in \cite{Gubser:2008px} that the scalar field acquires negative contributions to its effective mass squared which, at low temperatures, can be such that the mass drops below the BF bound over a large enough range to render the system unstable to the formation of scalar hair\footnote{We use the term ``hair'' in the spirit of the no hair theorems, \cite{Bekenstein:1996pn,Hertog:2006rr}, as in something that leaves an imprint at radial infinity.}.
This non-trivial scalar hair has a power law fall off at the boundary and the coefficient of this fall off can be interpreted, via the correspondence, as a condensate in the boundary theory.  Analysis of the boundary theory has shown it to exhibit many of the qualitative characteristics of a superconducting system.  There has been a great deal of recent work in this area, studying how the system behaves in a variety of different scenarios including altering the scalar mass and spacetime dimension, changing the gauge group and adding an external magnetic field \cite{Hartnoll:2008vx, Hartnoll:2008kx,Horowitz:2008bn,Gubser:2009cg,Horowitz:2009ij,Gubser:2008wv,Peeters:2009sr,Nakano:2008xc,Kanno:2010pq}.  Such systems have become known as holographic superconductors.  

The majority of these models are, so called, ``bottom up'' models\footnote{For an example of some ``top down'' approaches see \cite{Gubser:2009qm,Gauntlett:2009dn, Gauntlett:2009bh}.} where the gravity theory, often Einstein gravity, is, according the gauge/gravity correspondence, thought of as a low energy effective theory for some overarching string theory.  In \cite{Gregory:2009fj,Barclay:2010up,Gregory:2010yr,Pan:2009xa,Pan:2010at,Cai:2010cv,Brihaye:2010mr,Siani:2010uw,Jing:2010cx} the authors studied the stability of these systems to the inclusion of the Gauss-Bonnet (GB) invariant in the gravitational action which is believed to be the $\mathcal{O}(\alpha^\prime)$ correction to some low energy string theories \cite{Gross:1986mw,Metsaev:1987zx}. Thus, by studying this system the authors were studying the stability of the model to the inclusion of higher order corrections.  The findings of these papers showed that the qualitative features of the holographic superconductor were stable to higher order corrections but the details changed.  These papers however, used either only one particular mass to focus their analysis or confined their studies to the probe, or non backreacting, limit.  This has meant that the full panorama of the GB invariant on the backreacting superconductor has remained unknown.

This paper expands upon \cite{Barclay:2010up}, studying the fully backreacting holographic superconductor in GB gravity for a variety of scalar masses for positive and negative GB coupling constant $\alpha$.  In a recent paper, \cite{Buchel:2009sk}, it was suggested that causality constraints from hydrodynamics limit the GB coupling to $\alpha \in [-0.711,0.113]$, in this work, however, we permit the full range of $\alpha\in(-\infty,L^2/4]$ in our study for greater understanding of its effect.  We study the critical temperature of the superconducting system, both analytically and numerically for a variety of scalar masses.  We find that in a region of parameter space close to both the BF bound and the upper limit of $\alpha$, numerical analysis is possible even at large, super-planckian scales and that its effect is to increase the critical temperature, not reduce it, as is its tendency in other regions of parameter space.  The study of negative $\alpha$ shows $T_c$ to increase as $\alpha$ drops and the effect of backreaction is diminished as $\alpha$ gets large and negative.  

We then look at the zero temperature limit of this superconductor first showing analytically that there can be no regular superconducting solutions at zero temperature for systems with a tachyonic scalar field and place strict constraints on systems where the mass is positive.  Following \cite{Horowitz:2009ij} we relax the constraint of regularity for a system without a black hole, allowing logarithmically divergent terms.  We find that such solutions are incompatible with the concept of GB gravity being a perturbation of Einstein gravity.  In the absence of reliable numerical zero temperature solutions  we then use two analytic arguments to place bounds on the critical values of the constants of the theory about which the zero temperature phase transitions may occur.

We also study the effect that the GB coupling, backreaction and the mass have on the conductivity of the system, finding largely quantitative not qualitative alterations, the exception being the effect of backreaction on the appearance of quasi-normal modes in the conductivity for a system in the vicinity of the BF bound.  We find that increasing backreaction quickly removes the appearance of these quasi-normal modes within the temperature range that we are numerically able to study.

The paper is organized as follows: Section 2 provides a brief overview of the holographic superconductor in GB gravity and a presentation of the equations of motion that are to be solved.  In section 3 we study the nature of the condensate and in particular the critical temperature of the system for which we use both analytic and numerical results.  In section 4 we study the zero temperature limit.  In section 5 we study the conductivity of the system. Finally we conclude in section 6.


\section{The bulk}

In this section we review the set up introduced in \cite{Barclay:2010up}. This consisted of an Einstein Gauss-Bonnet (EGB) gravitational action coupled to a massive charged complex scalar field and a U(1) gauge field
\begin{align}\label{action}
S&=\frac{1}{2\kappa^2}\int d^5x \sqrt{-g} \left[
-R + \frac{12}{L^2} + \frac{\alpha}{2} \left(
R^{abcd}R_{abcd} -4R^{ab}R_{ab} + R^2
\right) \right]\nonumber\\
&\hspace{3.5mm}+\int d^5x\sqrt{-g}\left[
-\frac{1}{4}F^{ab}F_{ab}+|\nabla_a\psi -iqA_a\psi|^2
-V(|\psi|)
\right]\,,
\end{align}
where  $\kappa^2 = 8\pi G_5$ provides an explicit Planck scale in the system, $g$ is the determinant of the metric,  $R$, $R_{abcd}$ and $R_{ab}$ are the Ricci scalar, Riemann curvature and Ricci tensors respectively.  The negative cosmological constant, $-6/L^2$, has been written in terms of a length scale, $L$ and $\alpha\in(-\infty,L^2/4]$ is the GB coupling constant. $\bf{A}$ is the gauge field and $\psi$ is a scalar field with charge $q$.

In this paper we will study the minimal potential consisting of a single term, quadratic in $\psi$
\begin{align}
V(|\psi|)=m^2|\psi|^2\label{Pot1},
\end{align}
where $m$ is the mass of scalar field.

In order to examine holographic superconductivity we look
for charged, planar, black hole solutions\footnote{For studies of GB black holes in adS space see \cite{Boulware:1985wk,Cai:2001dz,Charmousis:2002rc}.} with or without nontrivial scalar hair. We do this by using the following metric and static field Ans\"atze
\begin{gather}
ds^2 = f(r)e^{2\nu(r)}dt^2 - \frac{dr^2}{f(r)} 
- \frac{r^2}{L_e^2}(dx^2+dy^2+dz^2),\label{metric}\\
A_a=\phi(r)\delta^0_a,\quad\quad\quad\quad\quad\quad \psi=\psi(r),
\end{gather}
where without loss of generality $\psi$ can be taken to be real.  $L_e$ is the effective asymptotic lengthscale of this space time given by
\begin{eqnarray}\label{Leff}
L^2_{\rm e}=\frac{L^2}{2}\left(1+\sqrt{1-\frac{4\alpha}{L^2}}\right) 
\to  \left\{
\begin{array}{rl}
\frac{L^2}{2}  \ , &  \quad {\rm for} \  \alpha \rightarrow \frac{L^2}{4} \\
L^2   \ , &  \quad {\rm for} \ \alpha \rightarrow 0 \\
\infty  \ , &  \quad {\rm for} \  \alpha \rightarrow -\infty.
\end{array}\right.
\end{eqnarray} 

The fully coupled system of gravity, gauge and scalar equations take the simple form\footnote{Note that there is an $ij$ component of the EGB equations but it is not independent of (\ref{nueq}) and (\ref{feq})}
\begin{align}
&\phi^{\prime\prime}+\left( \frac{3}{r}-\nu^\prime
\right)\phi^\prime -2q^2\frac{\psi^2}{f}\phi=0\,, \label{phieq}\\
&\psi^{\prime\prime}+\left( \frac{3}{r}+\nu^\prime+\frac{f^\prime}{f}
\right)\psi^\prime +\left(\frac{q^2\phi^2}{f^2e^{2\nu}}
-\frac{m^2}{f} \right)\psi=0\,, \label{psieq}\\
&\left(1-\frac{2\alpha f}{r^2} \right)\nu^\prime
=\frac{2\kappa^2}{3}r\left(
\psi^{\prime 2}+\frac{q^2\phi^2\psi^2}{f^2e^{2\nu}}\right) \label{nueq} ,\\
&\left(1-\frac{2\alpha f}{r^2} \right)f^\prime+\frac{2}{r}f
-\frac{4r}{L^2} =-\frac{2\kappa^2}{3}r\left[
\frac{\phi^{\prime2}}{2e^{2\nu}}+m^2\psi^2+
f\psi^{\prime2}+\frac{q^2\phi^2\psi^2}{fe^{2\nu}} \right]. \label{feq} 
\end{align}
where a prime denotes a derivative with respect to $r$.

In order to solve these equations we need to impose boundary conditions
at the horizon and the adS boundary.  The position of the horizon, $r_+$, is defined by $f(r_+)=0$.  Demanding regularity of the matter fields and metric at the horizon 
gives
\begin{eqnarray}
\phi(r_+)=0,\hspace{1cm}
\psi^\prime(r_+)=\frac{m^2}{f^\prime(r_+)}\psi(r_+) \ .
\end{eqnarray}

At the boundary we want the spacetime to asymptote adS in standard coordinates so we shall look for a solution with 
\begin{eqnarray}
&&\nu \to 0 \;\;\;,\;\;\;\;
f(r) \sim \frac{r^2}{L_e^2}\;\;\; {\rm as} \;\; r \to \infty\,.
\end{eqnarray}
Asymptotically the solutions of $\phi$ and $\psi$ are then found to be:
\begin{eqnarray}
\phi(r) \sim P - \frac{Q}{r^2}\,,\hspace{1cm} 
\psi(r) \sim \frac{C_{-}}{r^{\Delta_-}}+\frac{C_{+}}{r^{\Delta_+}}\,,
\label{r:boundary}
\end{eqnarray}
where $Q$ is the charge of the black hole (up to a factor of $4\pi$) and $\Delta_\pm=2\pm\sqrt{4+m^2L_e^2}$.   We choose to set $C_{-}=0$ and interpret, $ \langle {\cal O}_{\Delta_+} \rangle \equiv C_{+}$, where ${\cal O}_{\Delta_+}$ is a boundary operator with the conformal
dimension $\Delta_+$.  If $\Delta_\pm > 3$ the opposite choice of  $C_{+}=0$ and $ \langle {\cal O}_{\Delta_-} \rangle \equiv C_{-}$ does give normalizable solutions, but will not be considered in this work. An example of where such a choice is made for a system with Einstein gravity can be found in \cite{Horowitz:2008bn}.

This paper is concerned with the effect that varying the mass has on the superconductor.  We shall choose a sample of masses, greater or equal to that determined by the BF bound, $m^2=-4/L_e^2$, in order to observe the effect on the boundary theory.  Each choice of mass will be fixed with respect to the adS lengthscale, $L_e$, in order for the dimension of the boundary operator to remain constant with respect to variations in $\alpha$.

In the next section we solve (\ref{phieq}-\ref{feq}) numerically, reading this $1/r^{\Delta_+}$ fall off of the scalar field to obtain $ \langle {\cal O}_{\Delta_+} \rangle$ for a range of temperatures given by
 \begin{eqnarray}
 T = \frac{1}{4\pi} f' (r)e^{\nu(r)}\bigg|_{r=r_+}\ .
 \label{hawking}
 \end{eqnarray}
For numerical convenience we will use the scaling symmetries of the metric, \cite{Barclay:2010up}, to set $L=Q=q=1$.  With this rescaling $\kappa^2$ is the parameter used to vary the backreaction of the fields on the metric; if $\kappa^2=0$, referred to as the probe limit, the fields decouple from the metric entirely.


\section{The boundary}

We wish to study the effect that the inclusion of (GB) backreaction has on the holographic superconductor for the full range of masses. We begin by analysing the operator, $\langle {\cal O}_{\Delta_+} \rangle $, as a function of temperature, as seen in figure \ref{CondensatePlot}.  This plot shows that at high temperatures the expectation value of the boundary operator is zero which corresponds to the scalar field having a trivially zero profile in the bulk.  As the temperature drops below some critical temperature, $T_c$, the bulk scalar obtains a non-zero profile which, on the boundary, is interpreted as the operator `condensing' out of its vacuum.  The critical temperature is a feature particular to each system and can simply be read off from such plots.

\FIGURE{ 
\centering
\begingroup
  \makeatletter
  \providecommand\color[2][]{%
    \GenericError{(gnuplot) \space\space\space\@spaces}{%
      Package color not loaded in conjunction with
      terminal option `colourtext'%
    }{See the gnuplot documentation for explanation.%
    }{Either use 'blacktext' in gnuplot or load the package
      color.sty in LaTeX.}%
    \renewcommand\color[2][]{}%
  }%
  \providecommand\includegraphics[2][]{%
    \GenericError{(gnuplot) \space\space\space\@spaces}{%
      Package graphicx or graphics not loaded%
    }{See the gnuplot documentation for explanation.%
    }{The gnuplot epslatex terminal needs graphicx.sty or graphics.sty.}%
    \renewcommand\includegraphics[2][]{}%
  }%
  \providecommand\rotatebox[2]{#2}%
  \@ifundefined{ifGPcolor}{%
    \newif\ifGPcolor
    \GPcolortrue
  }{}%
  \@ifundefined{ifGPblacktext}{%
    \newif\ifGPblacktext
    \GPblacktexttrue
  }{}%
  \let\gplgaddtomacro\g@addto@macro
  \gdef\gplbacktext{}%
  \gdef\gplfronttext{}%
  \makeatother
  \ifGPblacktext
    \def\colorrgb#1{}%
    \def\colorgray#1{}%
  \else
    \ifGPcolor
      \def\colorrgb#1{\color[rgb]{#1}}%
      \def\colorgray#1{\color[gray]{#1}}%
      \expandafter\def\csname LTw\endcsname{\color{white}}%
      \expandafter\def\csname LTb\endcsname{\color{black}}%
      \expandafter\def\csname LTa\endcsname{\color{black}}%
      \expandafter\def\csname LT0\endcsname{\color[rgb]{1,0,0}}%
      \expandafter\def\csname LT1\endcsname{\color[rgb]{0,1,0}}%
      \expandafter\def\csname LT2\endcsname{\color[rgb]{0,0,1}}%
      \expandafter\def\csname LT3\endcsname{\color[rgb]{1,0,1}}%
      \expandafter\def\csname LT4\endcsname{\color[rgb]{0,1,1}}%
      \expandafter\def\csname LT5\endcsname{\color[rgb]{1,1,0}}%
      \expandafter\def\csname LT6\endcsname{\color[rgb]{0,0,0}}%
      \expandafter\def\csname LT7\endcsname{\color[rgb]{1,0.3,0}}%
      \expandafter\def\csname LT8\endcsname{\color[rgb]{0.5,0.5,0.5}}%
    \else
      \def\colorrgb#1{\color{black}}%
      \def\colorgray#1{\color[gray]{#1}}%
      \expandafter\def\csname LTw\endcsname{\color{white}}%
      \expandafter\def\csname LTb\endcsname{\color{black}}%
      \expandafter\def\csname LTa\endcsname{\color{black}}%
      \expandafter\def\csname LT0\endcsname{\color{black}}%
      \expandafter\def\csname LT1\endcsname{\color{black}}%
      \expandafter\def\csname LT2\endcsname{\color{black}}%
      \expandafter\def\csname LT3\endcsname{\color{black}}%
      \expandafter\def\csname LT4\endcsname{\color{black}}%
      \expandafter\def\csname LT5\endcsname{\color{black}}%
      \expandafter\def\csname LT6\endcsname{\color{black}}%
      \expandafter\def\csname LT7\endcsname{\color{black}}%
      \expandafter\def\csname LT8\endcsname{\color{black}}%
    \fi
  \fi
  \setlength{\unitlength}{0.0500bp}%
  \begin{picture}(6802.00,3968.00)%
    \gplgaddtomacro\gplbacktext{%
      \csname LTb\endcsname%
      \put(814,715){\makebox(0,0)[r]{\strut{}0}}%
      \put(814,1865){\makebox(0,0)[r]{\strut{}0.5}}%
      \put(814,3014){\makebox(0,0)[r]{\strut{}1}}%
      \put(946,484){\makebox(0,0){\strut{}0}}%
      \put(2038,484){\makebox(0,0){\strut{}0.05}}%
      \put(3130,484){\makebox(0,0){\strut{}0.1}}%
      \put(4221,484){\makebox(0,0){\strut{}0.15}}%
      \put(5270,484){\makebox(0,0){\strut{}$T_c$}}%
      \put(6405,484){\makebox(0,0){\strut{}0.25}}%
      \put(176,2203){\rotatebox{-270}{\makebox(0,0){\strut{}$\langle\mathcal{O}\rangle^{1/3} L$}}}%
      \put(3675,154){\makebox(0,0){\strut{}$T L$}}%
    }%
    \gplgaddtomacro\gplfronttext{%
    }%
    \gplbacktext
    \put(0,0){\includegraphics{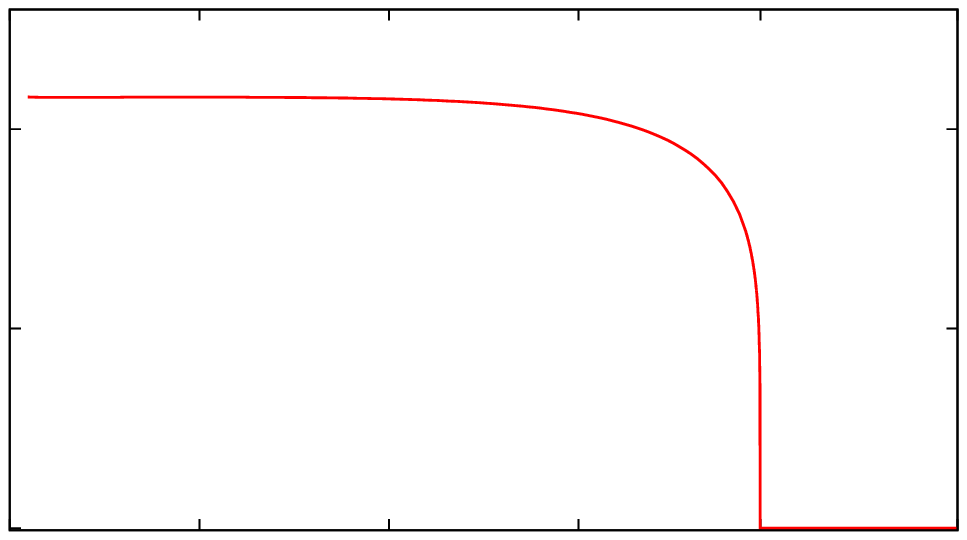}}%
    \gplfronttext
  \end{picture}%
\endgroup

 \caption{Plot of the condensate as a function of temperature; $\alpha=\kappa^2=0$, $m^2=-3/L_e^2$.}
   \label{CondensatePlot}
}

Varying $\alpha$, $\kappa^2$ and $m^2$ produces qualitatively similar plots to those of figure \ref{CondensatePlot} with the key differences being a variation in $T_c$.  As well as obtaining the exact value of $T_c$ from such numerically produced plots, a quicker, but rougher understanding can be obtained from an analytically generated lower bound on $T_c$ first introduced in \cite{Barclay:2010up}.   This bound is found by studying the scalar field equation in the vicinity of $T_c$. At temperatures just below $T_c$, the scalar field is small, $\psi\ll 1$, and the metric and gauge field will have the form
\begin{align}
\phi_0(r) &= \frac{Q}{r_+^2} \left ( 1 - \frac{r_+^2}{r^2} \right),\label{eq:GBAdSRNBlackHole1}\\
f_0(r) &= \frac{r^2}{2\alpha} \left [ 1 \pm \sqrt{1 - \frac{4\alpha}{L^2} \left ( 1 - \frac{r_+^4}{r^4} \right )+ \frac{8\alpha\kappa^2 Q^2}{3r^4 r_+^2} \left ( 1 -\frac{r_+^2}{r^2}\right )} \right],\label{eq:GBAdSRNBlackHole}
\end{align}
up to corrections of order ${\cal O} (\psi^2)$.  Thus,  in this region, the scalar field satisfies a linear equation, (\ref{psieq}), with $f$ and $\phi$ taking their background values. By letting $Y = r^3\psi$ and rearranging the field equation for $Y$ implies that {\it if} a solution exists, then the integral
\begin{align}
\int_{r_+}^\infty \frac{1}{r^3}  \left [\frac{\phi_0^2}{f_0} + \frac{3}{L_e^2} 
+ \frac{3f_0}{r^2} - \frac{3f_0'}{r} \right ]dr = - \int_{r_+}^\infty
\frac{f_0 Y^{\prime2}}{r^3Y^2}\, dr \leq 0
\label{eq:LowerBoundInt}
\end{align}
is negative. For much of parameter space this integral is negative at large $T$, and positive as $T\to0$, thus, observing where it changes sign provides a lower bound on $T_c$.  As is noted in \cite{Barclay:2010up}, the negativity of this integral does not imply existence of a solution to the linearised equation near $T_c$ but is simply a necessary condition on one if it exists. Therefore, any result obtained from the lower bound must be supported by numerically calculated values of $T_c$.  

Figures \ref{TcWithAlpha} to \ref{NegA} show both the analytic lower bound (as lines) and numerical values (as points) of $T_c$ for different values of $\alpha$, $\kappa^2$ and $m^2$.  Figure \ref{TcWithAlpha}  demonstrates the dependence of $T_c$ on $m^2$, focussing on $\alpha\geq0$.  It is possible to find superconducting solutions for $m^2>0$, indeed we found solutions up to a mass of $m^2\approx0.4$ for $\kappa^2=0$, reducing slightly as $\kappa^2$ was increased. The findings of \cite{Kim:2009kb} suggest that solutions exist at even larger values but that numerical solutions become difficult to obtain due to an intriguing ``warping'' of the space of permissible boundary conditions.  The solutions that we did obtain for a small positive mass were only marginally different to those of $m^2=0$ and so these plots have not been included.  In the plots that are shown we see that in the majority of parameter space  the effect of increasing backreaction is to reduce the value of $T_c$.  However, as $\alpha\to L^2/4$ and $m^2\to -4/L_e^2$ the effect of backreaction is reversed and actually increases $T_c$.  This can be explored in more detail by plotting $T_c$ as a function of $\kappa^2$, as seen in figure \ref{CondKappa}.  This plot clearly shows that in this very narrow region of parameter space the effect of backreaction can be to increase the critical temperature of the system substantially above its value in the probe limit.  The ability to reach super-planckian values of backreaction has been verified numerically up to $\kappa^2\approx150$.  It is also interesting to note that as one approaches this regime the lower bound on the critical temperature becomes significantly less accurate. 
\FIGURE{
\begin{tabular}{cc}
(a) $m^2=0$ & (b) $m^2=-1/L_e^2$\\
\includegraphics[width=7.25cm]{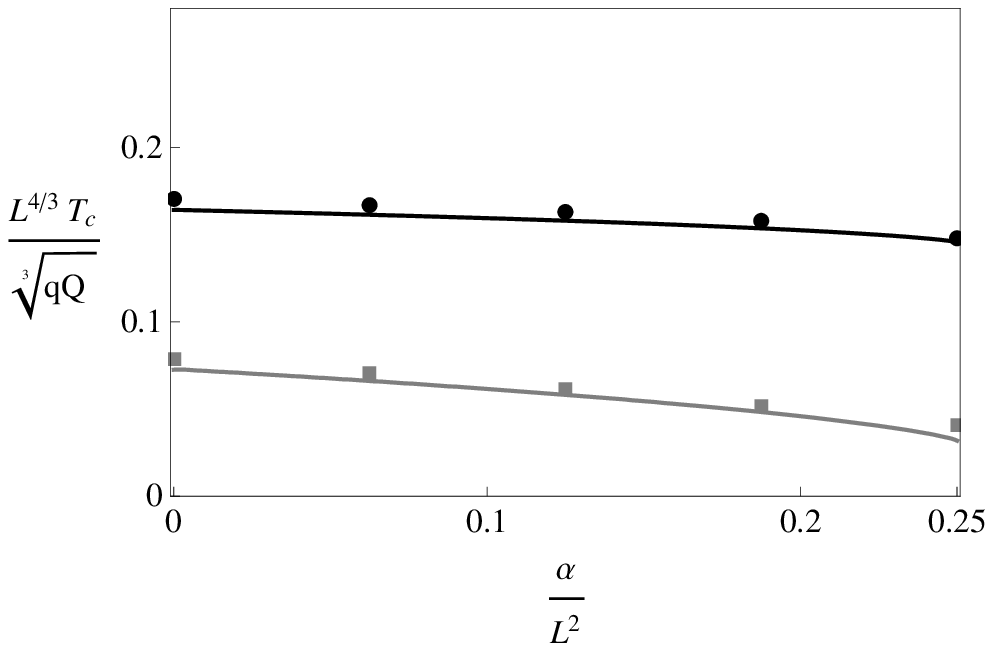} & 
\includegraphics[width=7.25cm]{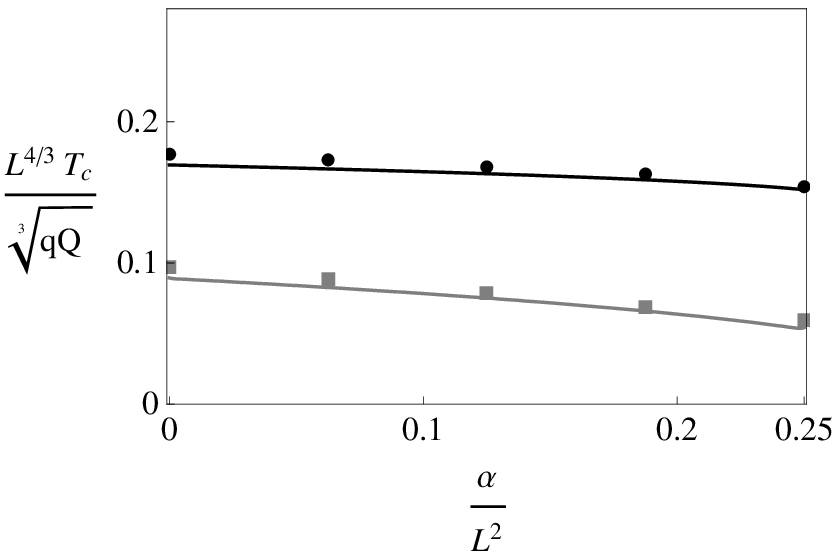} \\
(c) $m^2=-2/L_e^2$ & (d) $m^2=-3/L_e^2$ \\
\includegraphics[width=7.25cm]{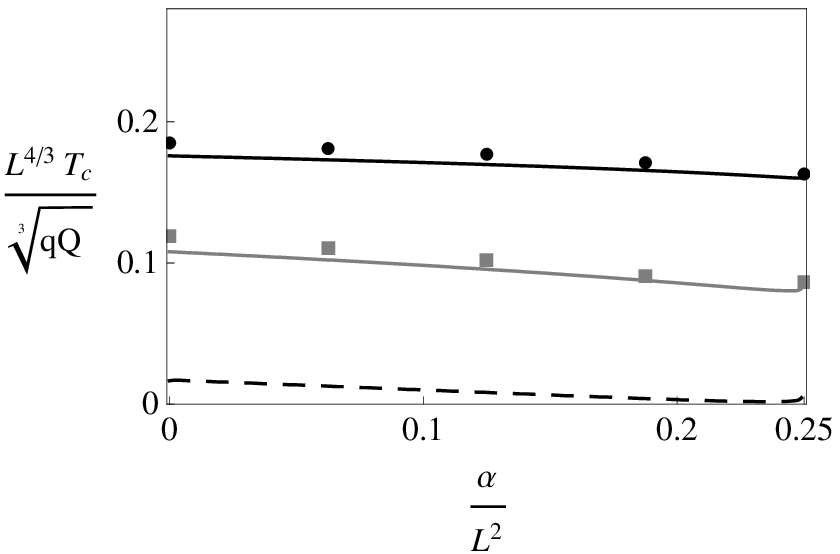} & 
\includegraphics[width=7.25cm]{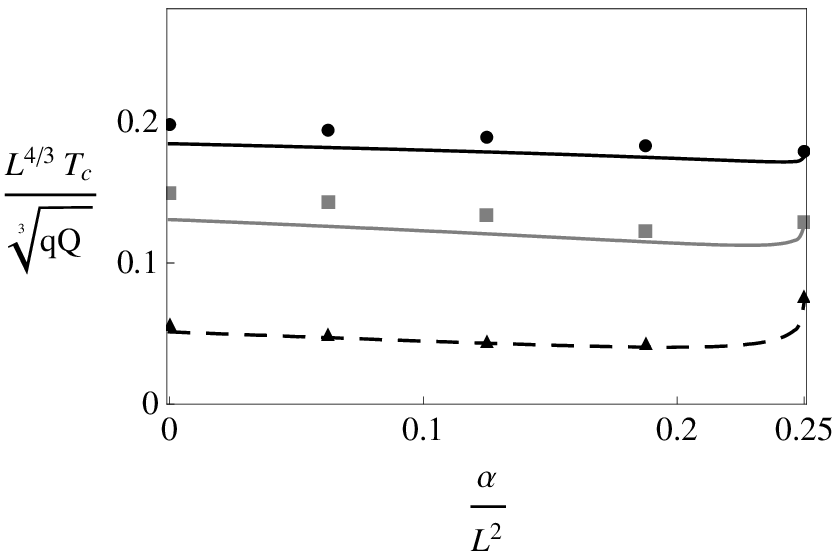} \\
(e) $m^2=-3.75/L_e^2$ & (f) $m^2=-4/L_e^2$\\
\includegraphics[width=7.25cm]{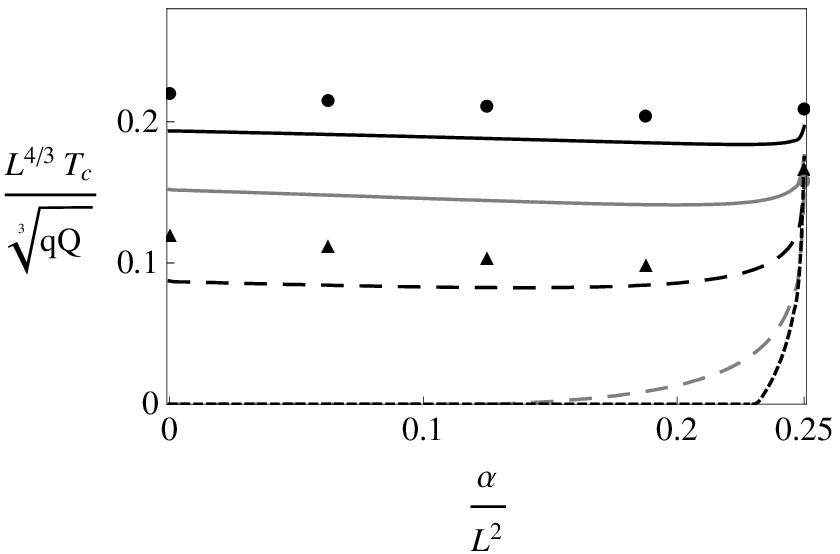} & 
\includegraphics[width=7.25cm]{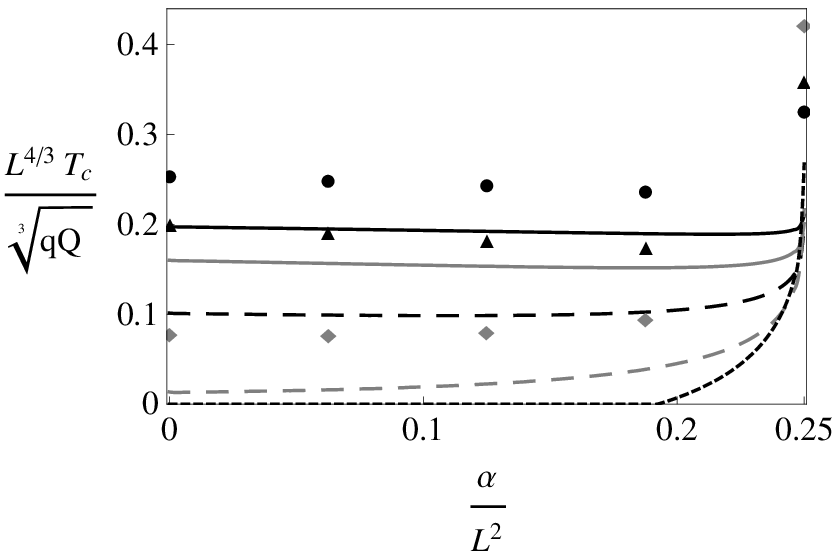} 
\end{tabular}
\caption{Plot of $T_c$ as a function of $\alpha$ for a variety of $\kappa^2$; Lines represent the analytic lower bound, and the points represent numerically obtained values. The solid black lines and circular points corresponds to $\kappa^2=0.0$, solid grey lines and square points to $\kappa^2=0.05$, black (large) dashed with triangular points to $\kappa^2=0.2$, grey (large) dashed and diamond points to $\kappa^2=1$ and black (small) dashed to $\kappa^2=5$.}
\label{TcWithAlpha}
}

\FIGURE{
   \centering
\includegraphics[width=9cm]{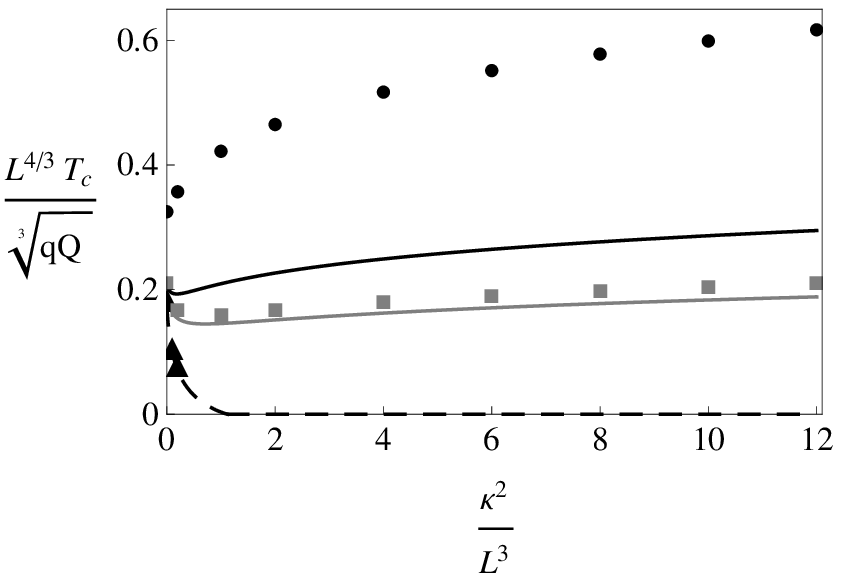}
\caption{Plot of $T_c$ as a function of $\kappa^2$ at $\alpha=0.24999$ for different masses; For each mass the analytic lower bounds are represented as lines and numerical values as points.  The black solid line with circular points corresponds to $m^2=-4/L_e^2$, grey with square points to $m^2=-3.75/L_e^2$ and the  black dashed line with triangular points to  $m^2=-3/L_e^2$.}
\label{CondKappa}
}

It is straightforward to extend this analysis to $\alpha<0$, as shown in figure \ref{NegA}.  In this regime the effect of altering the mass is less marked so one mass of $m^2=-2/L_e^2$ has been chosen as a representative sample.  These plots show that as $\alpha$ becomes more negative the critical temperature increases.  Whilst this increase becomes more and more gradual as $\alpha$ is reduced it appears that an arbitrarily large $T_c$ can be obtained by an appropriate choice of $\alpha$.   These plots also show that the effect of backreaction is, in all cases, to reduce $T_c$, but as $\alpha$ becomes large and negative its effect is diminished.    This can be understood by looking at the action, (\ref{action}).  In the Einstein limit the curvature of the spacetime scales with $\kappa$.  When $|\alpha|$ is large the higher order curvature terms dominate meaning the curvature scales as $\sqrt{\kappa}$ and thus the effect of backreaction on the spacetime is reduced.

\FIGURE{
\begin{tabular}{cc}
(a) & (b)  \\
\includegraphics[width=7.25cm]{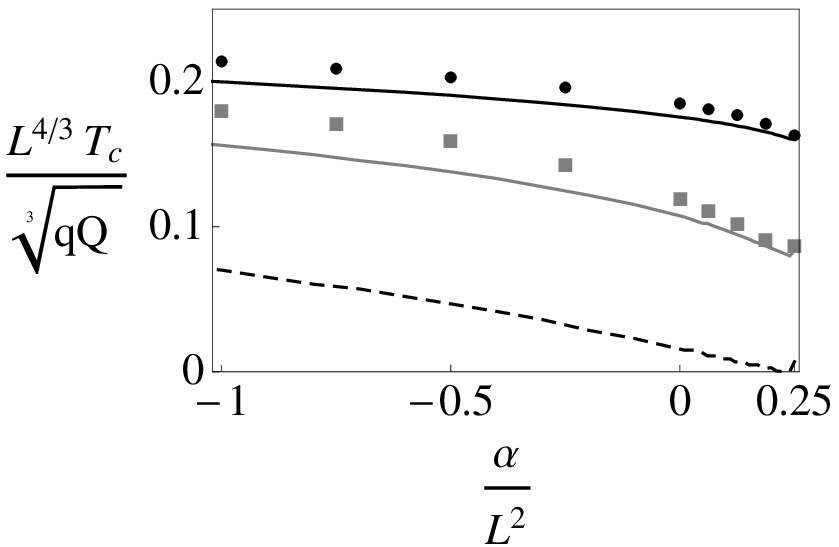} &
\includegraphics[width=7.25cm]{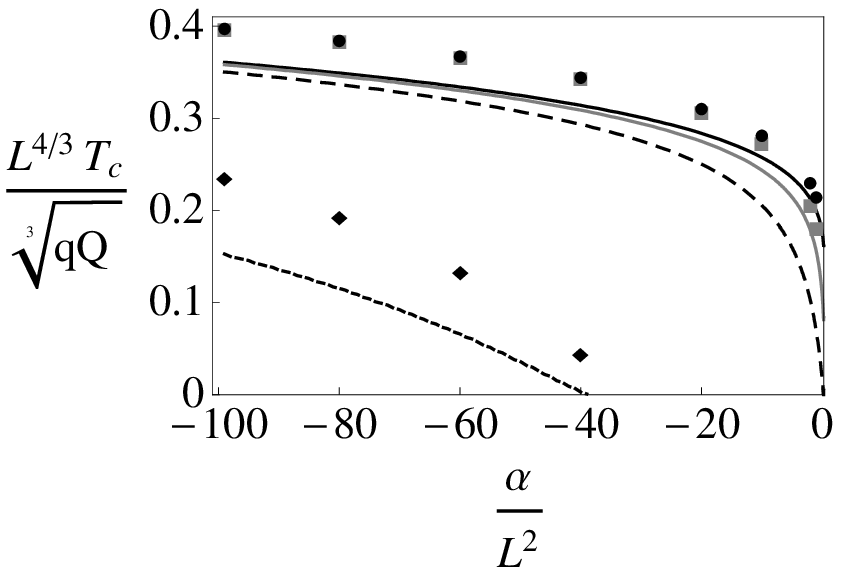}  
\end{tabular}
\caption{Plot of $T_c$ with $\alpha$ for $m^2=-2/L_e^2$.  (a) shows the region $\alpha\in[-1,0.25]$ and (b) shows the same plot but for $\alpha\in[-100,0.25]$.  The lines correspond to the lower bound, points to numerically obtained values of $T_c$.  The black solid lines (and circular points) correspond to $\kappa^2=0$; solid grey (and square points) to $\kappa^2=0.05$; dashed black to $\kappa^2=0.2$ and (smaller) dashed black (with diamond points) to $\kappa^2=5$. }
\label{NegA}
}

In an attempt to provide a clearer picture of the characteristics noted above we can use the analytically calculated lower bound on $T_c$ and scan through the parameter space available to generate the lower bound on a surface of $T_c$ as seen in figure \ref{LowerBoundSurface}. 
\FIGURE{
  (a) \qquad\qquad\qquad\qquad\qquad\qquad\qquad\qquad (b)  \\
  \includegraphics[scale=0.7125]{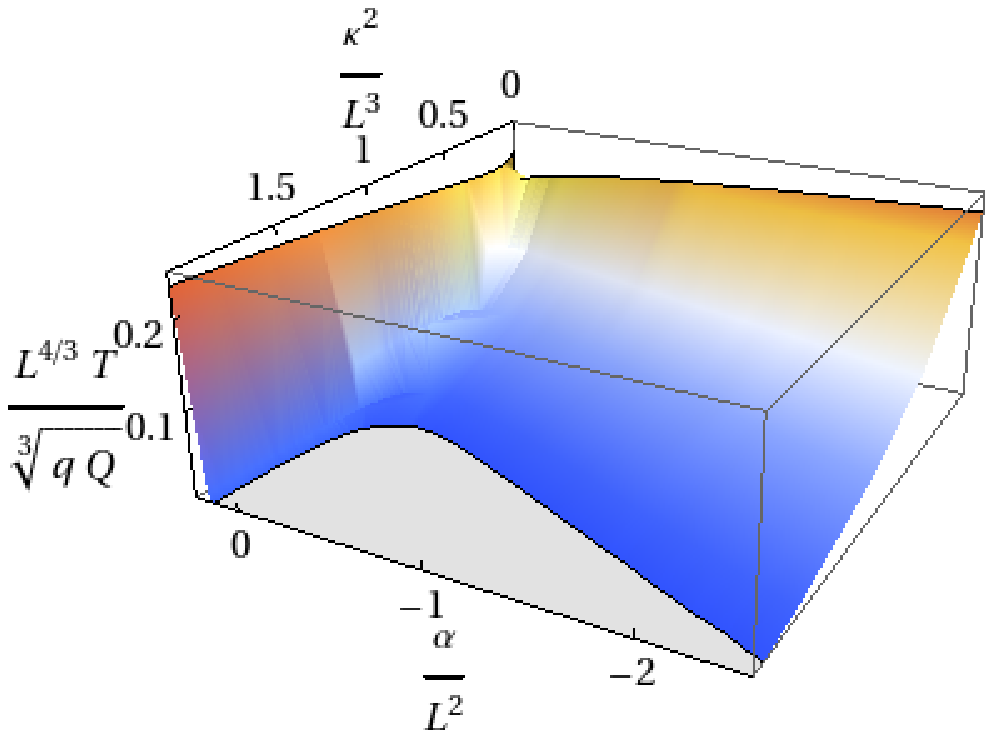} 
  \includegraphics[scale=0.7125]{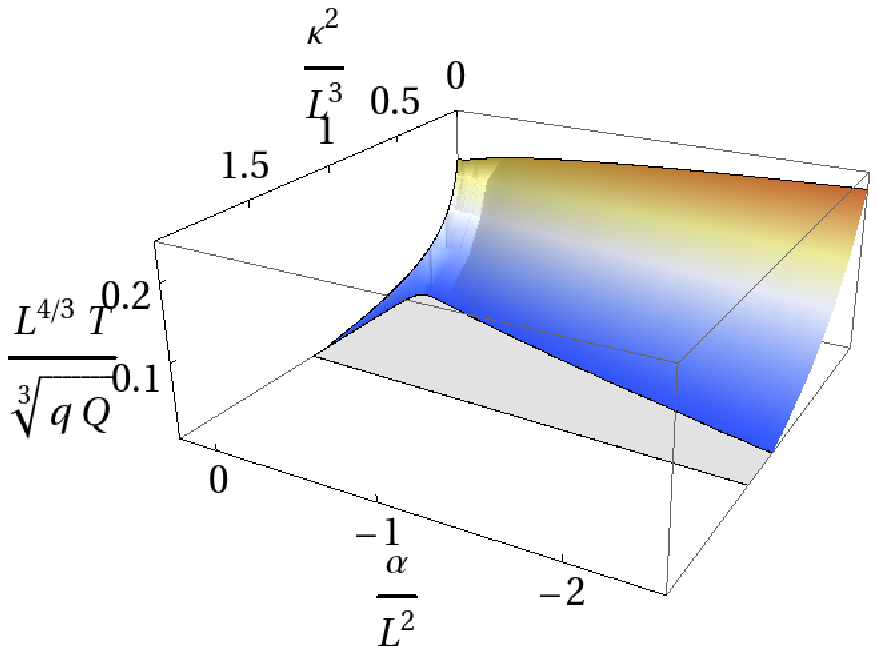} \\
\qquad\qquad(c) \\
  \includegraphics[scale=0.7125]{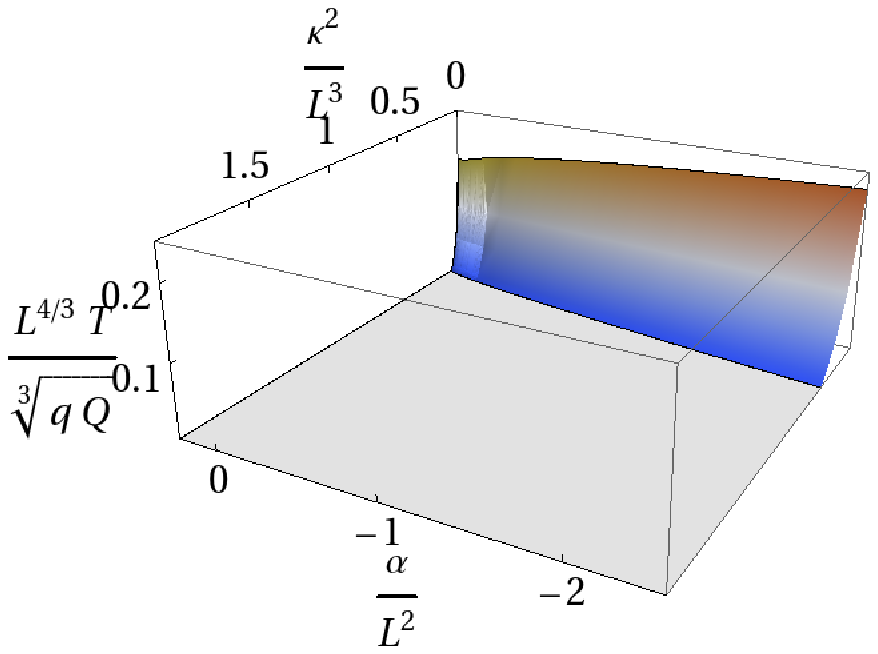} 
\caption{Plots (a), (b) and (c) show the surface of a lower bound on $T_c$ for $m^2=-4/L_e^2$, $-3/L_e^2$ and $0$ respectively.}
\label{LowerBoundSurface}
} Whilst these plots are only a lower bound to the true surface of $T_c$ they do exhibit some of the interesting characteristics of the system that have been supported by exact numerical results.  We see immediately how altering the mass of the scalar field dramatically alters the nature of this superconducting system.


\section{Zero Temperature Superconductors}

There is a great deal of interest in the zero temperature limit of these superconducting systems and in particular in the phase transitions that happen there.  Most phase transitions are triggered by the thermal fluctuations of the system but at zero temperature, where there are no thermal fluctuations, phase transitions are triggered by the quantum fluctuations associated with Heisenberg's uncertainty principle.  The critical points about which these zero temperature phase transitions occur are called quantum critical points (QCPs). It is thought that in certain regimes the effect of the QCP can extend to finite temperature giving rise to unusual physical phenomena.  For real superconducting systems it is impossible to reach the absolute zero temperature required to study these QCPs. However, this is not the case with our theoretical models leading to a great deal of recent activity in this direction.

For holographic superconductors the temperature of the boundary theory is governed by the temperature of the black hole in the bulk spacetime.  The temperature of a black hole in our system is given by 
\begin{align}
T=\frac{1}{4\pi}f^\prime(r)e^{\nu(r)}\bigg|_{r=r_+}.  \label{eq:HawkTemp}
\end{align}
In general the temperature of a black hole can approach zero in a variety of ways depending on the type of black hole.  For example, above the critical temperature of our system the black holes are simply Reissner-Nordstr\"om black holes in GB gravity.   Such black holes have
\begin{align}
f^\prime(r_+)=\frac{4r_+}{L^2}-\frac{4\kappa^2Q^2}{3r_+^5}
\end{align} which means that the mass and charge can balance such that the temperature goes to zero at finite $r_+$.  This is not the case for the uncharged Schwarzschild black hole that arises when $\kappa^2=0$.  This has $f^\prime(r_+)=4r_+/L^2$ and the zero temperature limit is approached when $r_+\rightarrow0$.  To study this limit of the holographic superconductor we must investigate the hairy black hole. A priori it is not immediately obvious how such black holes approach zero temperature; is it found at some finite $r_+$ or when $r_+\rightarrow0$.  From the numerical solutions calculated above it seems that the latter may be true since the temperature is reduced by reducing $r_+$ and within the range studied there have been no apparent zero temperature solutions at finite $r_+$.  However, it is numerically very difficult to approach $r_+=0$ from some finite value and it is possible that a zero temperature finite $r_+$ solution exists beyond the scope of the numerics.   In \cite{Horowitz:2009ij} the authors calculated numerical results for holographic superconductors in a regime where $r_+$ is precisely zero and the results of which, reassuringly, seemed to correspond the asymptote of their finite $r_+$ solutions.  However there are problems with this approach.  Firstly, that the $r_+=0$ spacetime is not continuously connected to the finite $r_+$ spacetime which introduces an element of uncertainty into the results. Also the results that they obtain are singular, which raises additional concerns.

We can attempt to find information about the true nature of the zero temperature superconductor with a little investigation of the field equations (\ref{phieq}) to (\ref{feq}).  In particular we will ask whether these equations permit the existence of zero temperature, regular solutions with a non-trivial scalar field?  We will show that this is not the case.  We extend the work of \cite{FernandezGracia:2009em} by showing that there are no regular zero temperature solutions, including those with $r_+=0$ for scalars with tachyonic masses.  We also address scalars with $m^2\geq0$.

We begin by imposing that our system be regular.  This will be true if the energy momentum tensor, $T_{\mu\nu}$, is non-singular in coordinates that are locally regular at the horizon, or indeed, at $r=0$ if there is no horizon.   Using Eddington-Finkelstein coordinates defined by $v=t+r^{*}$ and $r=\rho$ where $r^{*}$ is the tortoise coordinate defined by
\begin{align}
& dr^*=\frac{dr}{fe^{\nu}},
\end{align}
the following combinations of the energy momentum tensor must be regular
\begin{align}
T_{vv}&=T_{tt}=fe^\nu T^t_t,\label{reg3}\\
T_{v\rho}&=\frac{-T_{tt}}{fe^\nu}=-e^\nu T^t_t,\\
T_{\rho\rho}&=T_{rr}+\frac{T_{tt}}{f^2e^{2\nu}}=\frac{1}{f}(T^t_t-T_r^r). \label{reg}
\end{align}
(\ref{reg}) gives the most restrictive constraint, namely that
\begin{align}
\frac{\phi^2\psi^2}{f^2e^{2\nu}}+{\psi^\prime}^2 <\infty
\end{align} must be finite and hence each of the individual terms must also be finite.   We wish to assess whether the field equations  permit these constraints to hold for a non-trivial solution at zero temperature.  The field equations are unchanged by the coordinate transformation and we are free to use (\ref{phieq}) to (\ref{feq}) in our analysis.

Note that (\ref{nueq}), plus the regularity of $(T^0_0-T^r_r)/f$ implies that $\nu^\prime(r_+)$ is regular.  If $\nu^\prime(r_+)$ is regular then $\nu(r_+)$ is regular and $e^{\nu(r_+)}\neq0$.  Thus from the definition of the temperature of our black hole, (\ref{eq:HawkTemp}),  the requirement of zero temperature must imply that $f^\prime(r_+)=0$.

We shall now study what effect this constraint has on the scalar field equation
\begin{align}
f\psi^{\prime\prime}+\left(\frac{3}{r}+\nu^\prime+\frac{f^\prime}{f}\right)f \psi^\prime+\left( \frac{q^2\phi^2}{f e^{2\nu}}- m^2\right)\psi=0. \label{eq:PsiEqTimesF}
\end{align}
The terms containing $\psi^{\prime\prime}$ and $\psi^{\prime}$ go to zero at the horizon by the regularity of $\psi^\prime(r_+)$ and the fact that $f^\prime(r_+)=0$, thus the last term must also go to zero.  This implies that either  $\psi(r_+)=0$ or $m^2=\frac{q^2\phi^2}{fe^{2\nu}}$.  If $\psi(r_+)\neq0$ then, by (\ref{reg}),  $\frac{q^2\phi^2}{fe^{2\nu}}\to 0 $ which implies that $m^2=0$.  Our analysis does not rule out the existence of regular zero temperature solutions for this choice of mass. In fact, it seems likely that such solutions do exist in light of \cite{Horowitz:2009ij}, where similar solutions were found for a system in four dimensional Einstein gravity.  We leave the search for such solutions in this system to future research.

To investigate non-zero masses we consider $\psi(r_+)=0$.  If $m^2\leq0$ then all the leading order terms of (\ref{eq:PsiEqTimesF}) have the same sign and cannot balance irrespective of whether $r_+$ is finite or zero.  Thus there can be no regular, superconducting solutions at zero temperature for scalar fields with tachyonic masses.

Turning to $m^2>0$, where $\psi(r_+)=0$, it is possible to place strict constraints on these masses if solutions exist.  Using the field equation for $f$:
\begin{align}
 \left(1-\frac{2\alpha f}{r^2} \right)f^\prime+\frac{2}{r}f
-\frac{4r}{L^2} =-\frac{2\kappa^2}{3}r\left[
\frac{\phi^{\prime2}}{2e^{2\nu}}+m^2\psi^2+
f\psi^{\prime2}+\frac{q^2\phi^2\psi^2}{fe^{2\nu}} \right], \label{eq:Fequation2}
\end{align}
we see that if $\phi^\prime(r_+)=0$ then $r_+=0$ and $f(r)\sim r^2/L_e^2$ as $r\to0$.  From (\ref{eq:PsiEqTimesF}) we can then infer that $\phi(0)=0$ as otherwise $q^2\phi^2\psi/fe^{2\nu}$ would be the only term at leading order. Then from the field equation for $\phi$
\begin{align}
 \phi^{\prime\prime}+\phi^\prime\left( \frac{3}{r} -\nu^\prime \right)-2q^2\frac{\psi^2}{f}\phi=0, \label{eq:PhiEq2}
\end{align}
we see that the last term is sub-dominant and the remaining terms cannot cancel.

If $\phi^\prime(r_+)\neq0$, (\ref{eq:PhiEq2}) implies that $r_+\neq0$.  Then the leading and next to leading order terms of (\ref{eq:Fequation2}) give
\begin{align}
 \frac{\phi^{\prime 2}}{e^{2\nu_+}}&=\frac{12}{L^2\kappa^2}, & & f^{\prime\prime}_+=\frac{24}{L^2}.
\end{align}
By using these expressions in (\ref{eq:PsiEqTimesF}) we obtain an equation for the allowed masses at zero temperature
\begin{align}
 m^2=\frac{12}{L^2}(n^2+n)+\frac{q^2}{\kappa^2}, \label{eq:scalarMassZeroT}
\end{align}
where $n\geq1$ is the leading power of $(r-r_+)$ in an expansion of $\psi$ about $r=r_+$.

This expression shows that there can be no regular solutions for $0<m^2\leq24/L^2$.  Thus if positive mass solutions do exist they can only be found at very large $m^2$ and/or backreaction; substantially above the values for which finite temperature solutions have been found.  We also see that, unlike the finite temperature system, the ``allowed'' values of $m^2$ are directly related to $\kappa^2$. These observations suggest that this positive mass result may be spurious.

A key result of the above analysis is that the zero temperature limit of our superconducting systems with tachyonic 
 scalars is not regular.  We now wish to investigate this in a little more detail.  In \cite{Horowitz:2009ij} a zero temperature solution was presented in which the spacetime, with no black hole, possessed logarithmic divergences as $r\to0$.
 Such solutions can be found for our system in the Einstein limit but it becomes clear that they cannot be consistent with the idea of GB gravity as a perturbative expansion of Einstein gravity.  The reason is that the logarithmic divergences of the metric cause the curvature invariants, such as the Riemann tensor and Ricci scalar, to diverge at $r=0$.  Since the GB terms involve higher order combinations of these invariants than Einstein gravity this singular behaviour will immediately be dominated by the GB terms as $\alpha$ becomes non-zero.   If this is the case the concept of GB gravity being a perturbative correction to Einstein gravity is destroyed and the validity of such a solution must be questioned. 
 
 The manifestation of this problem on the fields themselves can be seen from a near horizon expansion.  Following \cite{Horowitz:2009ij},  for $\alpha=0$, one can find a set of boundary conditions consistent with the field equations
\begin{align}
&\psi=\sqrt{\frac{3}{\kappa^2}}(-\log r)^{1/2}+...,& & f=\frac{m^2}{2} r^2 \log r+..., \nonumber \\
&\phi=\phi_0 r^\beta (-\log r)^{1/2}+...,& & e^{2\nu}=K(\log r)^{-1} +...\label{logExpansion}
\end{align}
where  $\beta=-1+\sqrt{1-\frac{12q^2}{\kappa^2 m^2}}$ and $\phi_0/K$ is free parameter that can be used to tune the system. This Ansatz is consistent with the field equations provided $4q^2>-m^2\kappa^2$ and after integrating the fields out from the horizon one finds the asymptotic profiles to be consistent with (\ref{r:boundary}). Unlike in the four dimensional system we were unable to find an appropriate value of $\phi_0/K$ to remove the source of the boundary operator.  As a result these solutions do not strictly describe a holographic superconductor.  However, they are valid solutions of (\ref{phieq}) to (\ref{feq}) and can be used to demonstrate our point. 

 The problem for non-zero $\alpha$ arises because $\alpha$ appears in the equations of motion, (\ref{phieq}) to  (\ref{feq}), like $(1-\frac{2\alpha f}{r^2})$.  From (\ref{feq}) it is possible to show that if $f(r)=f_sr^s(-\log r)^t$ then $s\leq2$ which means that $f/r^2$ has at least a logarithmic singularity for $t>0$.  Since for $\alpha=0$, $t=1>0$ this means turning on $\alpha$ immediately incorporates new, singular behaviour at $r=0$ which destroys the perturbative relation between GB and Einstein gravity.


We have shown that there can be no regular solutions to our system at zero temperature, except possibly for massless or very massive scalars,  and we have also given cause for caution when considering non-regular solutions.  It is possible that consistent, non-regular, zero temperature solutions can be found that respect the relation between Einstein gravity and GB gravity but it seems unlikely.  However, we can still find out information about the nature of this system in the zero temperature regime in the absence of such solutions.  There are two analytic techniques which can provide bounds on the the critical values of the constants at the QCP.  The first bound is found by simply taking the zero temperature limit of the analytic lower bound argument for $T_c$.  In other words,  taking the zero temperature limit of plots such as those in figure \ref{LowerBoundSurface}.  These are shown as blue lines in figure \ref{T0Bounds}.

Another bound can be found by studying the stability of the vacuum solution to forming scalar hair.  Following \cite{Hartnoll:2009sz,Denef:2009tp} we study a
perturbation of $\psi$ about the vacuum solution by using the Ansatz
$\psi=\psi(r)e^{-i\omega t}$.  Assuming $\psi(r) \ll 1$ the effects of backreaction can be ignored (since its effects occur at $\mathcal{O}(\psi^2)$) and the scalar field equation becomes
\begin{align}\label{stabilityEqn}
  \psi^{\prime\prime}+\left(\frac{3}{r} +\frac{f^\prime}{f} 
\right)\psi^\prime+\left(\frac{\omega^2}{f^2}-\frac{2q\phi\omega}{f^2}-\frac{q^2\phi^2}{f^2}-\frac{m^2}{f}\right)\psi=0,
\end{align}
with $\phi$ and $f(r)$ taking their vacuum values (\ref{eq:GBAdSRNBlackHole1}) and (\ref{eq:GBAdSRNBlackHole}).

The system is unstable if the field equation shows this small perturbation to diverge.  This will be the case if there is a normalizable solution to (\ref{stabilityEqn}) with ingoing boundary conditions at the horizon such that $\omega$ has a positive imaginary part.  In general this equation can be solved numerically providing us with a bound on the critical values of the constants for general $T$.  However, we are just interested in the $T=0$ case for which an analytic expression can be obtained.

For zero temperature our (extremal) GB Reissner Nordstr\"om black hole has
\begin{align}
& \frac{r_+^6}{L^2}=\frac{\kappa^2Q^2}{3}
\label{extremalCondition}
\end{align} and
\begin{align}
&f(r) = \frac12
f_+^{\prime\prime}(r-r_+)^2+... \quad \quad\quad \text{with} \quad\quad
\quad  f_+^{\prime\prime}=\frac{24}{L^2}.\label{extremalCondition2}
\end{align}
It is then a simple exercise to expand (\ref{stabilityEqn}) in the vicinity of $r_+$.  It is suggested in  \cite{Hartnoll:2009sz,Denef:2009tp} that since we are concerned only with the onset of an instability it is sufficient to consider only the threshold case of $\omega=0$.  In this case we find the solution to the expanded field equation to be
\begin{align}
  &\psi\rightarrow c_1(r-r_+)^{\xi_+} +c_2(r-r_+)^{\xi_-},\\
&\xi_\pm=\frac12\left(1 \pm \sqrt{1-\frac{64 q^2Q^2}{r^6
{f^{\prime\prime}_+}^2}+\frac{8m^2}{f^{\prime\prime}_+}} \right).
\end{align}
If the expression inside the square root goes negative then $\psi$ will
turn imaginary and oscillate infinitely many times before reaching the horizon which, according to \cite{Denef:2009tp}, indicates an instability.  This provides a criterion determining the onset of an instability at extremality.
Using (\ref{extremalCondition}) and   (\ref{extremalCondition2}) if
\begin{align}\label{LowerBoundeq}
  3+m^2L^2<\frac{q^2Q^2L^4}{r_+^6 3}=\frac{q^2L^2}{\kappa^2}
\end{align}
the black hole is unstable to forming scalar hair\footnote{This same bound can be found by observing that the near horizon limit of the extremal RN black hole has topology $adS_2\times S_2$ which has a different, less negative, BF bound to the full spacetime.  Observing when the effective mass of the scalar field in the vicinity of the horizon is more negative then the $adS_2$ BF bound leads to precisely (\ref{LowerBoundeq}).}.

Figure \ref{T0Bounds} shows both this bound (as grey lines) and that taken from the zero temperature limit of the plots in figure \ref{LowerBoundSurface} (blue lines).
 \FIGURE{
(a)  \qquad\qquad\qquad\qquad\qquad\qquad\qquad\qquad (b)\\
  \includegraphics[width=7.2cm]{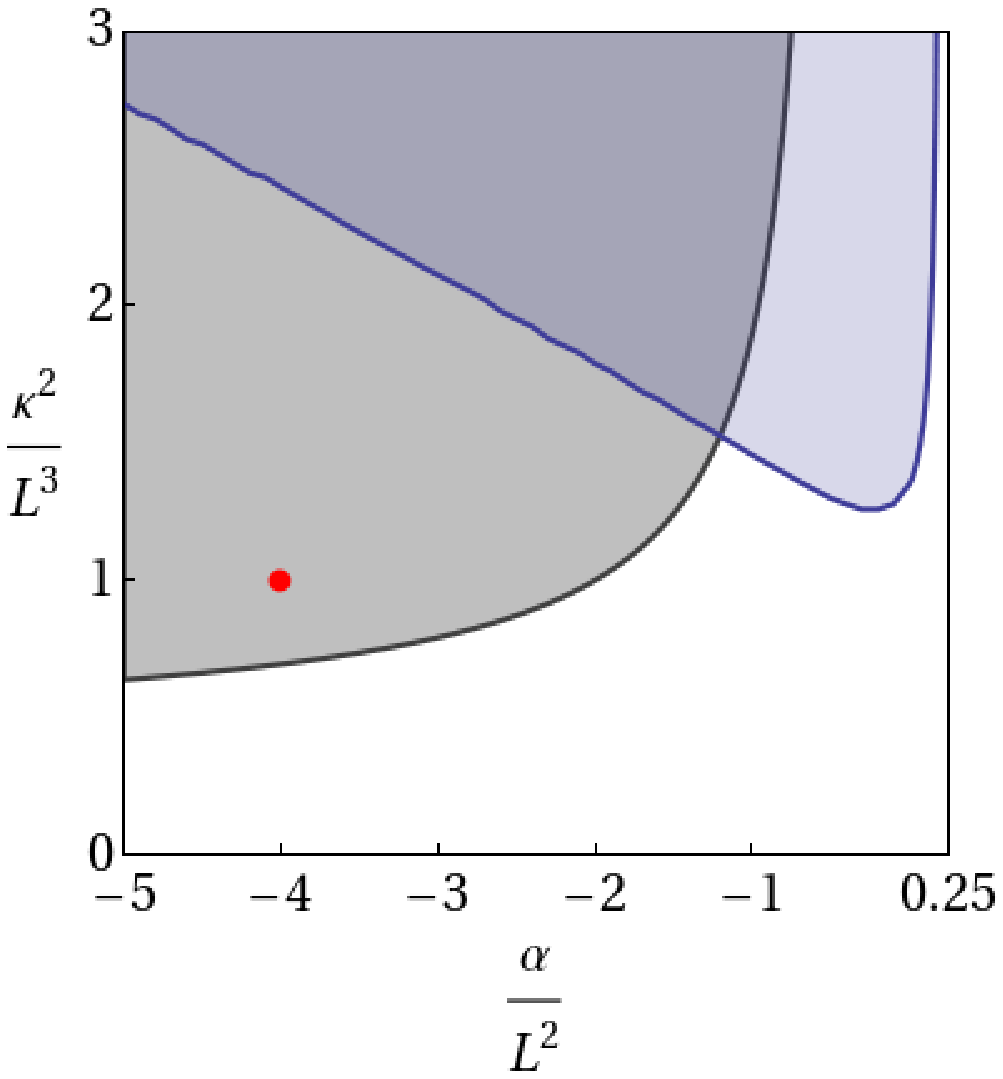} 
 \includegraphics[width=7.2cm]{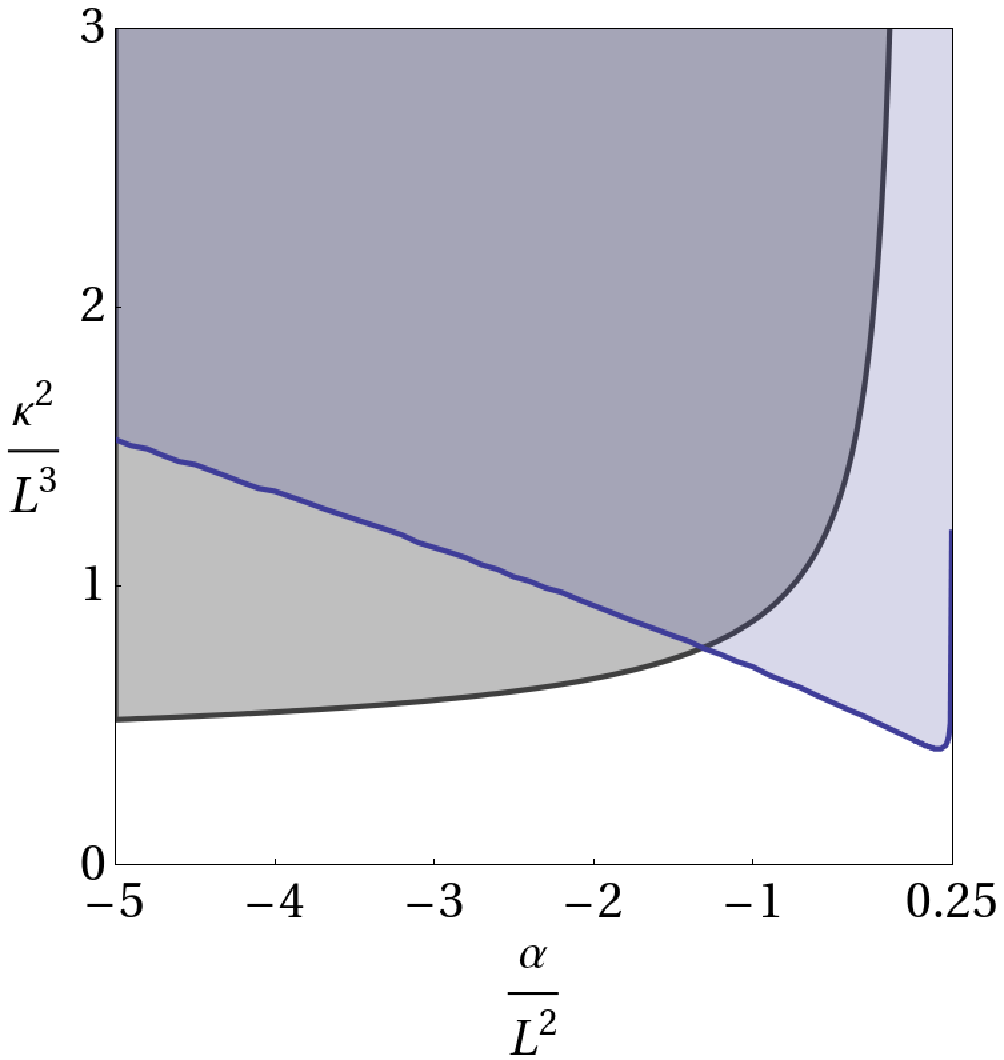}\\
  (c) \qquad\qquad\qquad\qquad\qquad\qquad\qquad\qquad (d)\\
    \includegraphics[width=7.2cm]{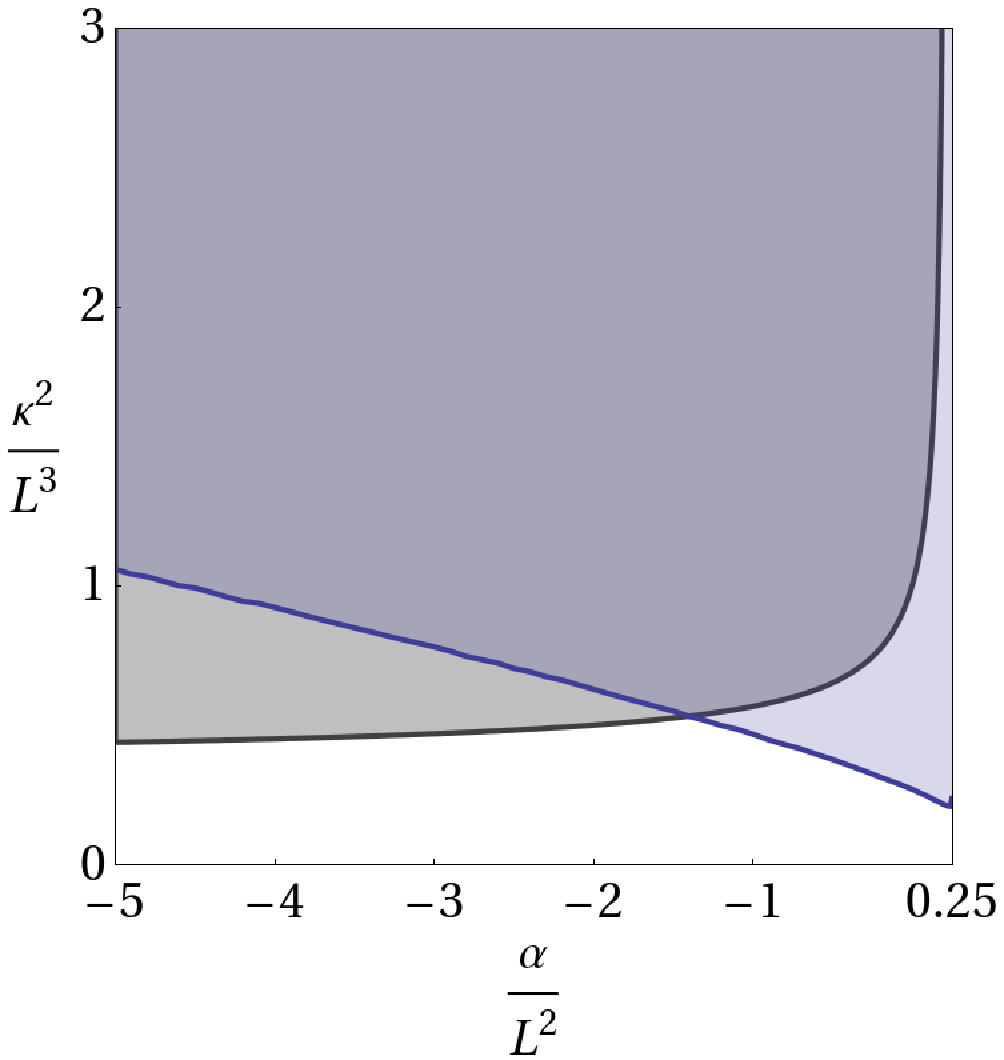} 
      \includegraphics[width=7.2cm]{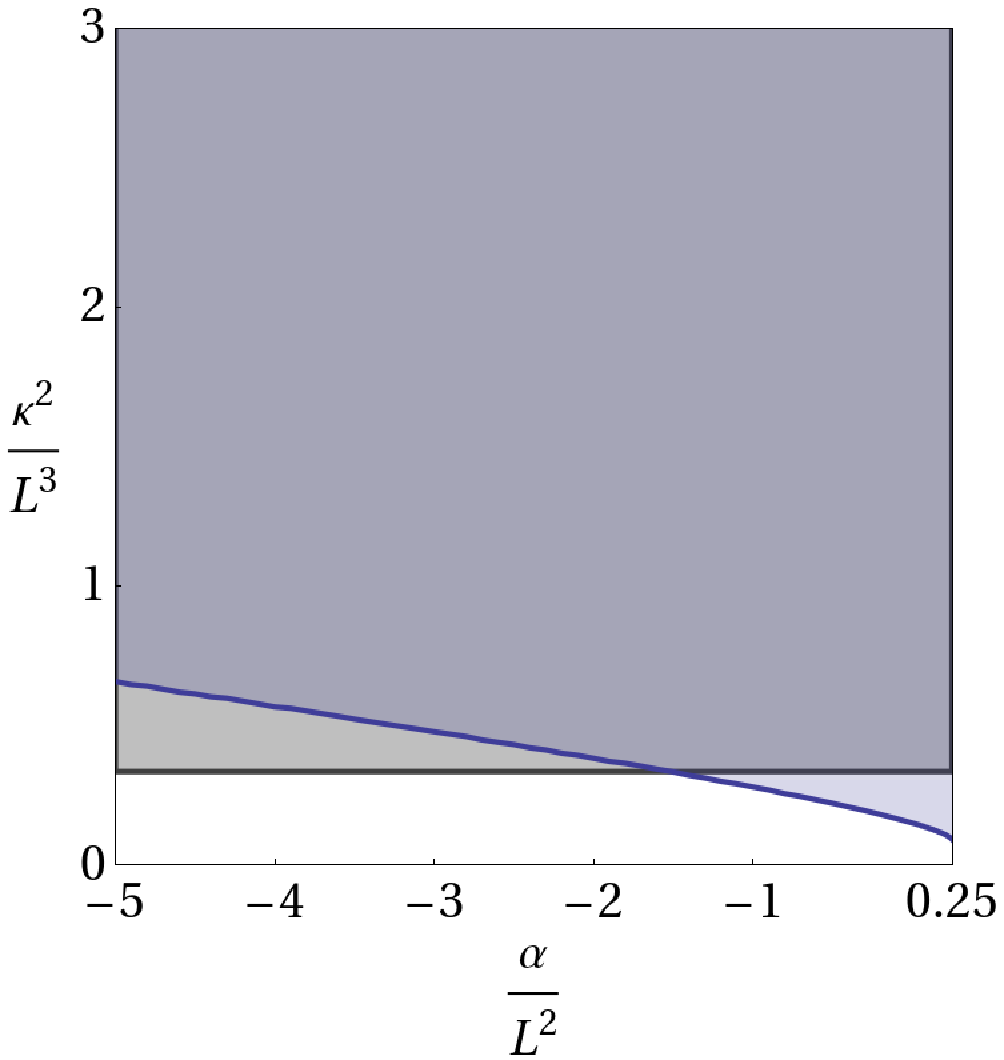}
\caption{Plots (a), (b), (c) and (d) show a bound, at  $T=0$, on the critical value of $\kappa^2$ as a function of $\alpha$ for $m^2=-4/L_e^2$, $-3/L_e^2$, $-2/L_e^2$ and $0$ respectively.  The region below each of the lines is the region where the system is unstable to forming scalar hair.  The blue lines were generated using the lower bound argument from (\ref{eq:LowerBoundInt}) whilst the gray lines were generated using (\ref{LowerBoundeq}).  The bounds continue to become less restrictive as $m^2$ increases above $0$.  The red point in (a) indicates a system with $m^2=-4/L_e^2$, $\alpha=-4$ and $\kappa^2=1$ for which the critical temperature was found to be $T_c=0.268$.}
   \label{T0Bounds}
}
The regions below each of the curves are the regions of parameter space for which the system is unstable to forming scalar hair, as indicated by each bound.  It was suggested in \cite{Hartnoll:2009sz,Denef:2009tp} that the grey lines are not simply a bound but actually indicate the location of the QCPs in the system.  Assuming that the true surface of critical temperature is continuous these plots immediately show that this cannot be the case as in each plot the two bounds cross.  This has been further verified by the calculation of a non-zero critical temperature for a system with $m^2=-4/L_e^2$, $\alpha=-4$ and $\kappa^2=1$ indicated by the red point in plot (a), which is outside the region of instability as indicated by bound (\ref{LowerBoundeq}).   
The correct way to see these curves is as complimenting lower bounds on the critical value of $\kappa^2$ as a function of $\alpha$, being aware that the true critical values could be some way above these combined bounds.  


The plots do however indicate that $m^2$ and $\alpha$ do have a significant effect on the zero temperature limit of the system with both bounds exhibiting the opening up of a region of superconductivity at large $\kappa^2$ as both $\alpha$ and $m^2$ approach their upper and lower bounds respectively.  This observation fully supports that of figure \ref{CondKappa} where numerically obtained values of $T_c$ were found at large, super-planckian backreaction.  From figure \ref{T0Bounds} (a) we see that this is unsurprising since in this region condensation must occur before the temperature drops to zero.  What, however, still remains unclear is why the critical temperature increases with backreaction here.   Another interesting observation that can be made from these plots is that there can be no QCPs in the absence of backreaction.

It is also interesting to see how these bounds relate to equation (\ref{eq:scalarMassZeroT}) which expresses the values that $m^2$ must take if regular, positive mass, superconducting solutions exist at zero temperature.  Inserting (\ref{eq:scalarMassZeroT}) into (\ref{LowerBoundeq}) shows that that systems with these masses can never be in the unstable region as indicated by bound (\ref{LowerBoundeq}) and it is only for large and negative $\alpha$ that they can in the unstable region indicated by (\ref{eq:LowerBoundInt}).  This does not prove that these solutions do not exist but indicates that their existence may be unlikely.


\section{Conductivity}

The conductivity of our boundary theory, $\sigma$, is calculated by studying perturbations of the gauge field, $A_\mu$. We set $A_i(t,r,x^i)=A(r)e^{-i\omega t}e_i$ to be our perturbation and solve Maxwell's equation,
\begin{equation} \label{CondEq}
 A^{\prime\prime}+\left(\frac{f^\prime}{f}+\nu^\prime+\frac{1}{r}\right)A^\prime
+\left[\frac{\omega^2}{f^2e^{2\nu}}-\frac{2}{f}q^2\psi^2
-\frac{2\kappa^2r^2\phi^{\prime2}}{fe^{2\nu}\left(r^2-2\alpha f\right)}
\right]A =0 \;,
\end{equation}
with the physically imposed boundary condition of only in going radiation at the horizon:
\begin{align}
A(r) \sim f(r)^{-i\frac{\omega}{4\pi T_+}} \ .
\label{ingoing}
\end{align}
Here, $T_+$ is the Hawking temperature defined at $r=r_+$.  In \cite{Horowitz:2009ij} a very elegant interpretation of the holographic conductivity in four spacetime dimensions is provided.  It involves the recasting of their gauge equation to the form of a one dimensional Schr\"odinger equation
\begin{align}
-A,_{zz}+V(z)A=\omega^2 A,\label{Shrodinger}
\end{align} where $z$ is a new radial parameter.  $\sigma$ is then interpreted as a combination of the reflection and transmission coefficients of a wave passing through the potential $V(z)$.  Viewing it in this way allowed intuition from quantum mechanics to be used to understand many key aspects of the conductivity of their system.  Unfortunately, due to the higher dimensionality of the system discussed in this paper, such a treatment has proven less straightforward.  Transforming (\ref{CondEq}) in to the form of (\ref{Shrodinger}) requires a change of radial coordinate to $dz=\frac{dr}{fe^{\nu}}$ followed by a change of variable of $A=r^{-\frac{1}{2}}\tilde{A}$.  Proper treatment of this system via the Schr\"odinger equation requires $\tilde{A}$ to be normalizable.   Since $A(r\to\infty )$ is finite, $\tilde{A}(r\to\infty)$ is infinite and hence non-normalizable.  Instead, to calculate the conductivity we shall follow \cite{Barclay:2010up} and expand the solutions to (\ref{CondEq}) in the vicinity of the adS boundary, $(r\rightarrow\infty)$ giving
\begin{align}
\label{genasoln}
A=a_0 + \frac{a_2}{r^2}
+\frac{a_0 L_e^4 \omega^2}{2r^2}
\log \frac{r}{L},
\end{align}
with $a_0$ and $a_2$ being integration constants that are used to calculate the conductivity:
\begin{align}\label{ConductivityEquation}
\sigma=\frac{2a_2}{i\omega L_e^4 a_0 } 
+\frac{i\omega}{2} - i\omega \log \left ( \frac{L_e}{L} \right) \ .
\end{align}
As was noted in \cite{Barclay:2010up} and \cite{Horowitz:2008bn} there exists an arbitrariness of scale in the logarithmic term in (\ref{genasoln}) introduced during the holographic renormalization process \cite{Skenderis:2002wp}.  As a result there exists an arbitrariness of scale in the imaginary part of $\sigma$.  We shall take advantage of this fact in our numerical calculations by choosing an appropriate renormalization scale in order to present the characteristics of the conductivity most clearly.

Plotting $\sigma/T_c$ as a function of $\omega/T_c$ shows a number of key characteristics of the superconductor
\FIGURE{
   \centering
\begingroup
  \makeatletter
  \providecommand\color[2][]{%
    \GenericError{(gnuplot) \space\space\space\@spaces}{%
      Package color not loaded in conjunction with
      terminal option `colourtext'%
    }{See the gnuplot documentation for explanation.%
    }{Either use 'blacktext' in gnuplot or load the package
      color.sty in LaTeX.}%
    \renewcommand\color[2][]{}%
  }%
  \providecommand\includegraphics[2][]{%
    \GenericError{(gnuplot) \space\space\space\@spaces}{%
      Package graphicx or graphics not loaded%
    }{See the gnuplot documentation for explanation.%
    }{The gnuplot epslatex terminal needs graphicx.sty or graphics.sty.}%
    \renewcommand\includegraphics[2][]{}%
  }%
  \providecommand\rotatebox[2]{#2}%
  \@ifundefined{ifGPcolor}{%
    \newif\ifGPcolor
    \GPcolortrue
  }{}%
  \@ifundefined{ifGPblacktext}{%
    \newif\ifGPblacktext
    \GPblacktexttrue
  }{}%
  \let\gplgaddtomacro\g@addto@macro
  \gdef\gplbacktext{}%
  \gdef\gplfronttext{}%
  \makeatother
  \ifGPblacktext
    \def\colorrgb#1{}%
    \def\colorgray#1{}%
  \else
    \ifGPcolor
      \def\colorrgb#1{\color[rgb]{#1}}%
      \def\colorgray#1{\color[gray]{#1}}%
      \expandafter\def\csname LTw\endcsname{\color{white}}%
      \expandafter\def\csname LTb\endcsname{\color{black}}%
      \expandafter\def\csname LTa\endcsname{\color{black}}%
      \expandafter\def\csname LT0\endcsname{\color[rgb]{1,0,0}}%
      \expandafter\def\csname LT1\endcsname{\color[rgb]{0,1,0}}%
      \expandafter\def\csname LT2\endcsname{\color[rgb]{0,0,1}}%
      \expandafter\def\csname LT3\endcsname{\color[rgb]{1,0,1}}%
      \expandafter\def\csname LT4\endcsname{\color[rgb]{0,1,1}}%
      \expandafter\def\csname LT5\endcsname{\color[rgb]{1,1,0}}%
      \expandafter\def\csname LT6\endcsname{\color[rgb]{0,0,0}}%
      \expandafter\def\csname LT7\endcsname{\color[rgb]{1,0.3,0}}%
      \expandafter\def\csname LT8\endcsname{\color[rgb]{0.5,0.5,0.5}}%
    \else
      \def\colorrgb#1{\color{black}}%
      \def\colorgray#1{\color[gray]{#1}}%
      \expandafter\def\csname LTw\endcsname{\color{white}}%
      \expandafter\def\csname LTb\endcsname{\color{black}}%
      \expandafter\def\csname LTa\endcsname{\color{black}}%
      \expandafter\def\csname LT0\endcsname{\color{black}}%
      \expandafter\def\csname LT1\endcsname{\color{black}}%
      \expandafter\def\csname LT2\endcsname{\color{black}}%
      \expandafter\def\csname LT3\endcsname{\color{black}}%
      \expandafter\def\csname LT4\endcsname{\color{black}}%
      \expandafter\def\csname LT5\endcsname{\color{black}}%
      \expandafter\def\csname LT6\endcsname{\color{black}}%
      \expandafter\def\csname LT7\endcsname{\color{black}}%
      \expandafter\def\csname LT8\endcsname{\color{black}}%
    \fi
  \fi
  \setlength{\unitlength}{0.0500bp}%
  \begin{picture}(6802.00,4534.00)%
    \gplgaddtomacro\gplbacktext{%
      \csname LTb\endcsname%
      \put(814,874){\makebox(0,0)[r]{\strut{}-20}}%
      \put(814,1553){\makebox(0,0)[r]{\strut{}0}}%
      \put(814,2232){\makebox(0,0)[r]{\strut{}20}}%
      \put(814,2911){\makebox(0,0)[r]{\strut{}40}}%
      \put(814,3590){\makebox(0,0)[r]{\strut{}60}}%
      \put(814,4269){\makebox(0,0)[r]{\strut{}80}}%
      \put(946,484){\makebox(0,0){\strut{}0}}%
      \put(2290,484){\makebox(0,0){\strut{}5}}%
      \put(3662,484){\makebox(0,0){\strut{}10}}%
      \put(5033,484){\makebox(0,0){\strut{}15}}%
      \put(6405,484){\makebox(0,0){\strut{}20}}%
      \put(176,2486){\rotatebox{-270}{\makebox(0,0){\strut{}$\sigma/T_c$}}}%
      \put(3675,154){\makebox(0,0){\strut{}$\omega/T_c$}}%
    }%
    \gplgaddtomacro\gplfronttext{%
    }%
    \gplbacktext
    \put(0,0){\includegraphics{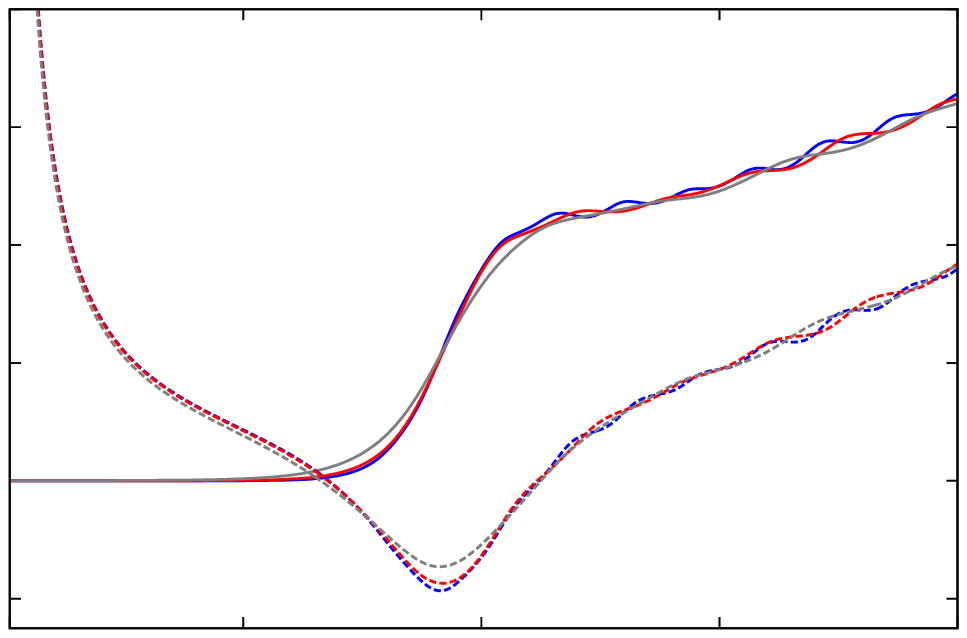}}%
    \gplfronttext
  \end{picture}%
\endgroup

   \caption{Conductivity: A plot showing the real (solid lines) and imaginary (dashed lines) parts of the conductivity, $\sigma/T_c$, as a function of $\omega/T_c$ for $m^2=-2/L_e^2$, $\alpha=0.125$, $\kappa^2=0$ at a variety of temperatures.  The grey, red and blue lines correspond to temperatures of $50\%$, $35\%$ and $25\%$ of the critical temperature respectively. The small oscillations at larger $\omega$ are a numerical artefact.}
   \label{Cond-Diff-Temp-away}
} as seen in figure \ref{Cond-Diff-Temp-away}.  This plot shows the real and imaginary parts of the conductivity for a boundary theory at $m^2=-2/L_e^2$, $\alpha=0.125$ and $\kappa^2=0$.  The first thing to note is a pole in $\text{Im}(\sigma)$ at $\omega=0$.  This, by the Kramers-Kronig relations which follow from causality, indicate the existence of a Dirac delta function in $\text{Re}(\sigma)$ at $\omega=0$.  This delta function, that cannot be picked up numerically, represents the infinite conductivity of the system.  The plot also clearly shows the presence of a step in  $\text{Re}(\sigma)$ coinciding with the global minimum of $\text{Im}(\sigma)$ which is interpreted as an energy gap in the superconductor.  Following \cite{Horowitz:2008bn} we let the value of this minimum be $\omega_g$, the value of the energy/frequency gap.  This plot also shows the effect that temperature has on the conductivity with the grey, red and blue lines corresponding to the measuring of $\sigma$ at $50\%$, $35\%$ and $25\%$ of the critical temperature respectively.  This shows that reducing the temperature alters the plot only slightly; making the step and dip sharper and more pronounced, but does not change the value of $\omega_g$.  Accessing lower temperatures has proved numerically very difficult. The absence of a reliable zero temperature solution means we can cast no light on what happens at as $T\rightarrow 0$.
 
 \FIGURE{
\begingroup
  \makeatletter
  \providecommand\color[2][]{%
    \GenericError{(gnuplot) \space\space\space\@spaces}{%
      Package color not loaded in conjunction with
      terminal option `colourtext'%
    }{See the gnuplot documentation for explanation.%
    }{Either use 'blacktext' in gnuplot or load the package
      color.sty in LaTeX.}%
    \renewcommand\color[2][]{}%
  }%
  \providecommand\includegraphics[2][]{%
    \GenericError{(gnuplot) \space\space\space\@spaces}{%
      Package graphicx or graphics not loaded%
    }{See the gnuplot documentation for explanation.%
    }{The gnuplot epslatex terminal needs graphicx.sty or graphics.sty.}%
    \renewcommand\includegraphics[2][]{}%
  }%
  \providecommand\rotatebox[2]{#2}%
  \@ifundefined{ifGPcolor}{%
    \newif\ifGPcolor
    \GPcolortrue
  }{}%
  \@ifundefined{ifGPblacktext}{%
    \newif\ifGPblacktext
    \GPblacktexttrue
  }{}%
  \let\gplgaddtomacro\g@addto@macro
  \gdef\gplbacktext{}%
  \gdef\gplfronttext{}%
  \makeatother
  \ifGPblacktext
    \def\colorrgb#1{}%
    \def\colorgray#1{}%
  \else
    \ifGPcolor
      \def\colorrgb#1{\color[rgb]{#1}}%
      \def\colorgray#1{\color[gray]{#1}}%
      \expandafter\def\csname LTw\endcsname{\color{white}}%
      \expandafter\def\csname LTb\endcsname{\color{black}}%
      \expandafter\def\csname LTa\endcsname{\color{black}}%
      \expandafter\def\csname LT0\endcsname{\color[rgb]{1,0,0}}%
      \expandafter\def\csname LT1\endcsname{\color[rgb]{0,1,0}}%
      \expandafter\def\csname LT2\endcsname{\color[rgb]{0,0,1}}%
      \expandafter\def\csname LT3\endcsname{\color[rgb]{1,0,1}}%
      \expandafter\def\csname LT4\endcsname{\color[rgb]{0,1,1}}%
      \expandafter\def\csname LT5\endcsname{\color[rgb]{1,1,0}}%
      \expandafter\def\csname LT6\endcsname{\color[rgb]{0,0,0}}%
      \expandafter\def\csname LT7\endcsname{\color[rgb]{1,0.3,0}}%
      \expandafter\def\csname LT8\endcsname{\color[rgb]{0.5,0.5,0.5}}%
    \else
      \def\colorrgb#1{\color{black}}%
      \def\colorgray#1{\color[gray]{#1}}%
      \expandafter\def\csname LTw\endcsname{\color{white}}%
      \expandafter\def\csname LTb\endcsname{\color{black}}%
      \expandafter\def\csname LTa\endcsname{\color{black}}%
      \expandafter\def\csname LT0\endcsname{\color{black}}%
      \expandafter\def\csname LT1\endcsname{\color{black}}%
      \expandafter\def\csname LT2\endcsname{\color{black}}%
      \expandafter\def\csname LT3\endcsname{\color{black}}%
      \expandafter\def\csname LT4\endcsname{\color{black}}%
      \expandafter\def\csname LT5\endcsname{\color{black}}%
      \expandafter\def\csname LT6\endcsname{\color{black}}%
      \expandafter\def\csname LT7\endcsname{\color{black}}%
      \expandafter\def\csname LT8\endcsname{\color{black}}%
    \fi
  \fi
  \setlength{\unitlength}{0.0500bp}%
  \begin{picture}(4534.00,3400.00)%
    \gplgaddtomacro\gplbacktext{%
      \csname LTb\endcsname%
      \put(814,716){\makebox(0,0)[r]{\strut{}0}}%
      \put(814,1926){\makebox(0,0)[r]{\strut{}50}}%
      \put(814,3135){\makebox(0,0)[r]{\strut{}100}}%
      \put(946,484){\makebox(0,0){\strut{}0}}%
      \put(1744,484){\makebox(0,0){\strut{}5}}%
      \put(2542,484){\makebox(0,0){\strut{}10}}%
      \put(3339,484){\makebox(0,0){\strut{}15}}%
      \put(4137,484){\makebox(0,0){\strut{}20}}%
      \put(176,1919){\rotatebox{-270}{\makebox(0,0){\strut{}$\sigma/T_c$}}}%
      \put(2541,154){\makebox(0,0){\strut{}$\omega/T_c$}}%
    }%
    \gplgaddtomacro\gplfronttext{%
    }%
    \gplbacktext
    \put(0,0){\includegraphics{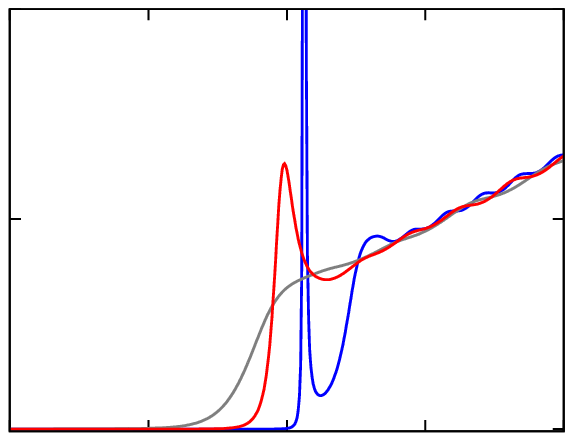}}%
    \gplfronttext
  \end{picture}%
\endgroup
\!\!\!\!\!\!\!
\begingroup
  \makeatletter
  \providecommand\color[2][]{%
    \GenericError{(gnuplot) \space\space\space\@spaces}{%
      Package color not loaded in conjunction with
      terminal option `colourtext'%
    }{See the gnuplot documentation for explanation.%
    }{Either use 'blacktext' in gnuplot or load the package
      color.sty in LaTeX.}%
    \renewcommand\color[2][]{}%
  }%
  \providecommand\includegraphics[2][]{%
    \GenericError{(gnuplot) \space\space\space\@spaces}{%
      Package graphicx or graphics not loaded%
    }{See the gnuplot documentation for explanation.%
    }{The gnuplot epslatex terminal needs graphicx.sty or graphics.sty.}%
    \renewcommand\includegraphics[2][]{}%
  }%
  \providecommand\rotatebox[2]{#2}%
  \@ifundefined{ifGPcolor}{%
    \newif\ifGPcolor
    \GPcolortrue
  }{}%
  \@ifundefined{ifGPblacktext}{%
    \newif\ifGPblacktext
    \GPblacktexttrue
  }{}%
  \let\gplgaddtomacro\g@addto@macro
  \gdef\gplbacktext{}%
  \gdef\gplfronttext{}%
  \makeatother
  \ifGPblacktext
    \def\colorrgb#1{}%
    \def\colorgray#1{}%
  \else
    \ifGPcolor
      \def\colorrgb#1{\color[rgb]{#1}}%
      \def\colorgray#1{\color[gray]{#1}}%
      \expandafter\def\csname LTw\endcsname{\color{white}}%
      \expandafter\def\csname LTb\endcsname{\color{black}}%
      \expandafter\def\csname LTa\endcsname{\color{black}}%
      \expandafter\def\csname LT0\endcsname{\color[rgb]{1,0,0}}%
      \expandafter\def\csname LT1\endcsname{\color[rgb]{0,1,0}}%
      \expandafter\def\csname LT2\endcsname{\color[rgb]{0,0,1}}%
      \expandafter\def\csname LT3\endcsname{\color[rgb]{1,0,1}}%
      \expandafter\def\csname LT4\endcsname{\color[rgb]{0,1,1}}%
      \expandafter\def\csname LT5\endcsname{\color[rgb]{1,1,0}}%
      \expandafter\def\csname LT6\endcsname{\color[rgb]{0,0,0}}%
      \expandafter\def\csname LT7\endcsname{\color[rgb]{1,0.3,0}}%
      \expandafter\def\csname LT8\endcsname{\color[rgb]{0.5,0.5,0.5}}%
    \else
      \def\colorrgb#1{\color{black}}%
      \def\colorgray#1{\color[gray]{#1}}%
      \expandafter\def\csname LTw\endcsname{\color{white}}%
      \expandafter\def\csname LTb\endcsname{\color{black}}%
      \expandafter\def\csname LTa\endcsname{\color{black}}%
      \expandafter\def\csname LT0\endcsname{\color{black}}%
      \expandafter\def\csname LT1\endcsname{\color{black}}%
      \expandafter\def\csname LT2\endcsname{\color{black}}%
      \expandafter\def\csname LT3\endcsname{\color{black}}%
      \expandafter\def\csname LT4\endcsname{\color{black}}%
      \expandafter\def\csname LT5\endcsname{\color{black}}%
      \expandafter\def\csname LT6\endcsname{\color{black}}%
      \expandafter\def\csname LT7\endcsname{\color{black}}%
      \expandafter\def\csname LT8\endcsname{\color{black}}%
    \fi
  \fi
  \setlength{\unitlength}{0.0500bp}%
  \begin{picture}(4534.00,3400.00)%
    \gplgaddtomacro\gplbacktext{%
      \csname LTb\endcsname%
      \put(814,704){\makebox(0,0)[r]{\strut{}-50}}%
      \put(814,1639){\makebox(0,0)[r]{\strut{}0}}%
      \put(814,2574){\makebox(0,0)[r]{\strut{}50}}%
      \put(946,484){\makebox(0,0){\strut{}0}}%
      \put(1738,484){\makebox(0,0){\strut{}5}}%
      \put(2538,484){\makebox(0,0){\strut{}10}}%
      \put(3337,484){\makebox(0,0){\strut{}15}}%
      \put(4137,484){\makebox(0,0){\strut{}20}}%
      \put(176,1919){\rotatebox{-270}{\makebox(0,0){\strut{}$\sigma/T_c$}}}%
      \put(2541,154){\makebox(0,0){\strut{}$\omega/T_c$}}%
    }%
    \gplgaddtomacro\gplfronttext{%
    }%
    \gplbacktext
    \put(0,0){\includegraphics{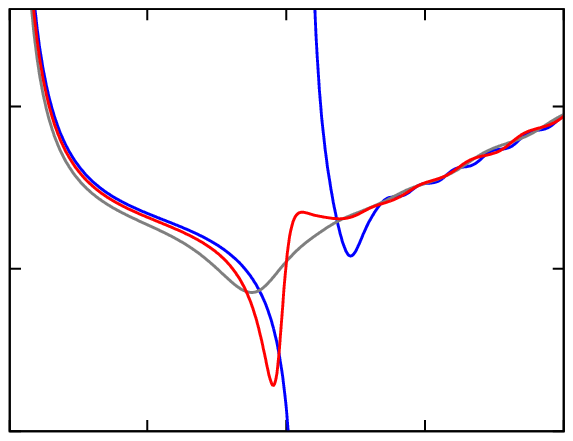}}%
    \gplfronttext
  \end{picture}%
\endgroup

  \caption{Conductivity: Plots showing the real (left) and imaginary (right) parts of the conductivity, $\sigma$, as a function of $\omega/T_c$ for $m^2=-4/L_e^2$, $\alpha=0.125$, $\kappa^2=0$ at a variety of temperatures.  The grey, red and blue lines correspond to temperatures of $50\%$, $35\%$ and $25\%$ of the critical temperature respectively.}
   \label{Cond-Diff-Temp-BF}
}

As we approach the BF bound the plot behaves quite differently, as can been seen in figure  \ref{Cond-Diff-Temp-BF}. Now we see that lowering the temperature does dramatically alter the plot.  At $T=0.5 T_c$ the plots looks very similar to that of figure \ref{Cond-Diff-Temp-away}, but as the temperature drops the step and dip shift to higher $\omega$, developing distinct peaks which turn into poles.  These poles are interpreted as quasi-normal frequencies that have moved to the real axis from elsewhere in the complex plane \cite{Horowitz:2008bn,Horowitz:2009ij, Brattan:2010pq}.  As the temperature drops further more poles appear at higher values of $\omega/T_c$ (not shown).  It is suggested in \cite{Siopsis:2010pi} that in the probe limit of Einstein gravity, as $T\rightarrow0$ the number of these poles diverges.  Since such low temperature analysis is outside the scope of this paper, we can shed no light on whether or not this occurs away from the Einstein limit.  

We are interested in observing the effect that varying $\alpha$, $\kappa^2$ and $m^2$ has on these phenomena.  We will begin by looking at the first case; away from the BF bound where temperature dependent effects are less prominent.  In \cite{Barclay:2010up} the authors studied the effect of $\alpha$, $\kappa^2$ for $m^2=-3/L_e^2$.  They found that increasing $\alpha$ above the Einstein limit increased the value of $\omega_g$ and made the step and dip more pronounced.  The effect of increasing $\kappa^2$ was to smooth out the features of the plot but not affecting the value of $\omega_g$, that is until the smoothing removes the presence of the hard gap\footnote{ i.e $\text{Re}(\sigma)$ no longer is zero for small $\omega$.}, at least within the temperature range studied.  Studying the conductivity for larger masses we see very similar results with quantitative differences rather than qualitative.  The key information has been captured on a plot of $\omega_g$ against $T_c$ as seen in figure \ref{OmegaVsTc}.    
\FIGURE{
\begin{tabular}{cc}
\includegraphics[width=7.25cm]{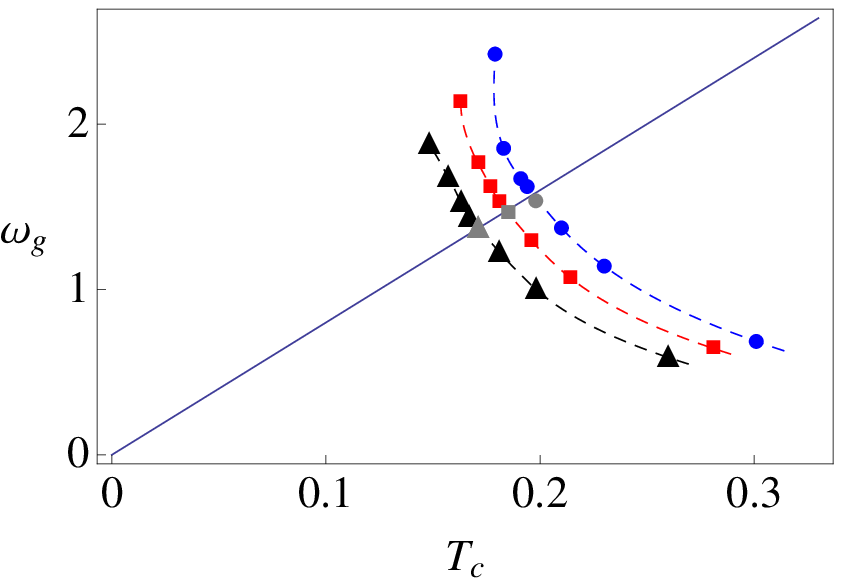}  &
\includegraphics[width=7.25cm]{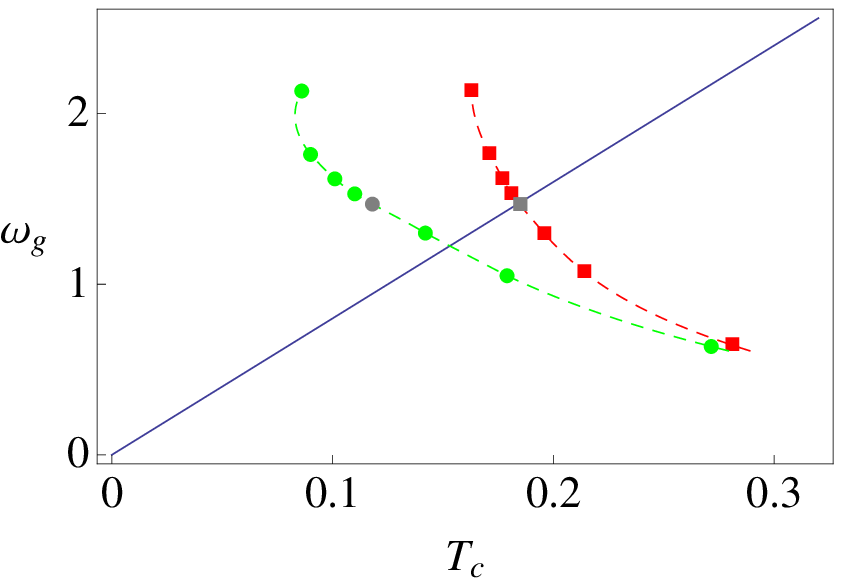}
\end{tabular}
\caption{ The left plot shows $\omega_g$ against $T_c$ for $\kappa^2=0$ and $m^2=0$, black (triangular) points; $m^2=-2/L_e^2$, red (square) points and $m^2=-3/L_e^2$, blue (circular) points.  The right plot shows $\omega_g$ against $T_c$ for  $m^2=-2/L_e^2$ for $\kappa^2=0$ red (square) points and $\kappa^2=0.05$ green (circular) points.  In both plots from top to bottom the points correspond to $\alpha=0.24999$, $0.1875$, $0.125$, $0.0625$, $0$, $-0.25$, $-1$, $-10$, with the grey points corresponding to Einstein gravity.  The dashed lines have been added to guide the eye.  The straight line corresponds to $\omega_g=8T_c$.}
\label{OmegaVsTc}
}

The grey points in left plot in figure \ref{OmegaVsTc} correspond to the probe, Einstein limit of the superconductor.  One can see that for the range of masses presented, the points all fall close to the line $\omega_g=8T_c$.  This observation contributed ammunition to the speculation, \cite{Horowitz:2008bn}, that this may be a universal relation.  This plot shows that such a relation is unstable to higher curvature corrections as found in \cite{Barclay:2010up}.  

The plot shows that increasing $\alpha$ increases $\omega_g$ and largely reduces $T_c$, except for very close to $\alpha=1/4$.  This has the effect of moving the point decidedly off the line.  Decreasing the mass from $m^2=0$ increases $\omega_g$ and $T_c$ with the greatest differences occurring towards the upper bound of $\alpha$ where variations in $T_c$ are more pronounced.  The right hand plot shows the effect of backreaction.  Increasing $\kappa^2$ has very little effect on $\omega_g$ with the majority of the effect coming from the reduction in $T_c$.  As $\alpha$ gets large and negative the points converge corresponding to the diminished effect of backreaction in this regime that was noted above.  We were unable to extend these curves to much larger negative coupling as the numerical artefacts that arise in our calculation of the conductivity began to obscure the key features of the plots.  However, since the  calculation of the condensate seems possible for arbitrarily large negative $\alpha$, one might expect these curves to continue towards the axis without ever reaching it.

We now turn our attention to systems at the BF bound, and in particular what effect $\alpha$ and $\kappa^2$ have on the development of the quasi-normal modes.  Figure \ref{QuasinormalvaryAlpha} shows $\text{Re}(\sigma)$, measured at $T=T_c/4$, for $m^2=-4/L_e^2$, $\kappa^2=0$ for a variety of values of $\alpha$.  
\FIGURE{
   \centering
\begingroup
  \makeatletter
  \providecommand\color[2][]{%
    \GenericError{(gnuplot) \space\space\space\@spaces}{%
      Package color not loaded in conjunction with
      terminal option `colourtext'%
    }{See the gnuplot documentation for explanation.%
    }{Either use 'blacktext' in gnuplot or load the package
      color.sty in LaTeX.}%
    \renewcommand\color[2][]{}%
  }%
  \providecommand\includegraphics[2][]{%
    \GenericError{(gnuplot) \space\space\space\@spaces}{%
      Package graphicx or graphics not loaded%
    }{See the gnuplot documentation for explanation.%
    }{The gnuplot epslatex terminal needs graphicx.sty or graphics.sty.}%
    \renewcommand\includegraphics[2][]{}%
  }%
  \providecommand\rotatebox[2]{#2}%
  \@ifundefined{ifGPcolor}{%
    \newif\ifGPcolor
    \GPcolortrue
  }{}%
  \@ifundefined{ifGPblacktext}{%
    \newif\ifGPblacktext
    \GPblacktexttrue
  }{}%
  \let\gplgaddtomacro\g@addto@macro
  \gdef\gplbacktext{}%
  \gdef\gplfronttext{}%
  \makeatother
  \ifGPblacktext
    \def\colorrgb#1{}%
    \def\colorgray#1{}%
  \else
    \ifGPcolor
      \def\colorrgb#1{\color[rgb]{#1}}%
      \def\colorgray#1{\color[gray]{#1}}%
      \expandafter\def\csname LTw\endcsname{\color{white}}%
      \expandafter\def\csname LTb\endcsname{\color{black}}%
      \expandafter\def\csname LTa\endcsname{\color{black}}%
      \expandafter\def\csname LT0\endcsname{\color[rgb]{1,0,0}}%
      \expandafter\def\csname LT1\endcsname{\color[rgb]{0,1,0}}%
      \expandafter\def\csname LT2\endcsname{\color[rgb]{0,0,1}}%
      \expandafter\def\csname LT3\endcsname{\color[rgb]{1,0,1}}%
      \expandafter\def\csname LT4\endcsname{\color[rgb]{0,1,1}}%
      \expandafter\def\csname LT5\endcsname{\color[rgb]{1,1,0}}%
      \expandafter\def\csname LT6\endcsname{\color[rgb]{0,0,0}}%
      \expandafter\def\csname LT7\endcsname{\color[rgb]{1,0.3,0}}%
      \expandafter\def\csname LT8\endcsname{\color[rgb]{0.5,0.5,0.5}}%
    \else
      \def\colorrgb#1{\color{black}}%
      \def\colorgray#1{\color[gray]{#1}}%
      \expandafter\def\csname LTw\endcsname{\color{white}}%
      \expandafter\def\csname LTb\endcsname{\color{black}}%
      \expandafter\def\csname LTa\endcsname{\color{black}}%
      \expandafter\def\csname LT0\endcsname{\color{black}}%
      \expandafter\def\csname LT1\endcsname{\color{black}}%
      \expandafter\def\csname LT2\endcsname{\color{black}}%
      \expandafter\def\csname LT3\endcsname{\color{black}}%
      \expandafter\def\csname LT4\endcsname{\color{black}}%
      \expandafter\def\csname LT5\endcsname{\color{black}}%
      \expandafter\def\csname LT6\endcsname{\color{black}}%
      \expandafter\def\csname LT7\endcsname{\color{black}}%
      \expandafter\def\csname LT8\endcsname{\color{black}}%
    \fi
  \fi
  \setlength{\unitlength}{0.0500bp}%
  \begin{picture}(6802.00,4534.00)%
    \gplgaddtomacro\gplbacktext{%
      \csname LTb\endcsname%
      \put(814,704){\makebox(0,0)[r]{\strut{}0}}%
      \put(814,1892){\makebox(0,0)[r]{\strut{}50}}%
      \put(814,3081){\makebox(0,0)[r]{\strut{}100}}%
      \put(814,4269){\makebox(0,0)[r]{\strut{}150}}%
      \put(946,484){\makebox(0,0){\strut{}0}}%
      \put(2760,484){\makebox(0,0){\strut{}10}}%
      \put(4582,484){\makebox(0,0){\strut{}20}}%
      \put(6405,484){\makebox(0,0){\strut{}30}}%
      \put(176,2486){\rotatebox{-270}{\makebox(0,0){\strut{}$\sigma/T_c$}}}%
      \put(3675,154){\makebox(0,0){\strut{}$\omega/T_c$}}%
    }%
    \gplgaddtomacro\gplfronttext{%
    }%
    \gplbacktext
    \put(0,0){\includegraphics{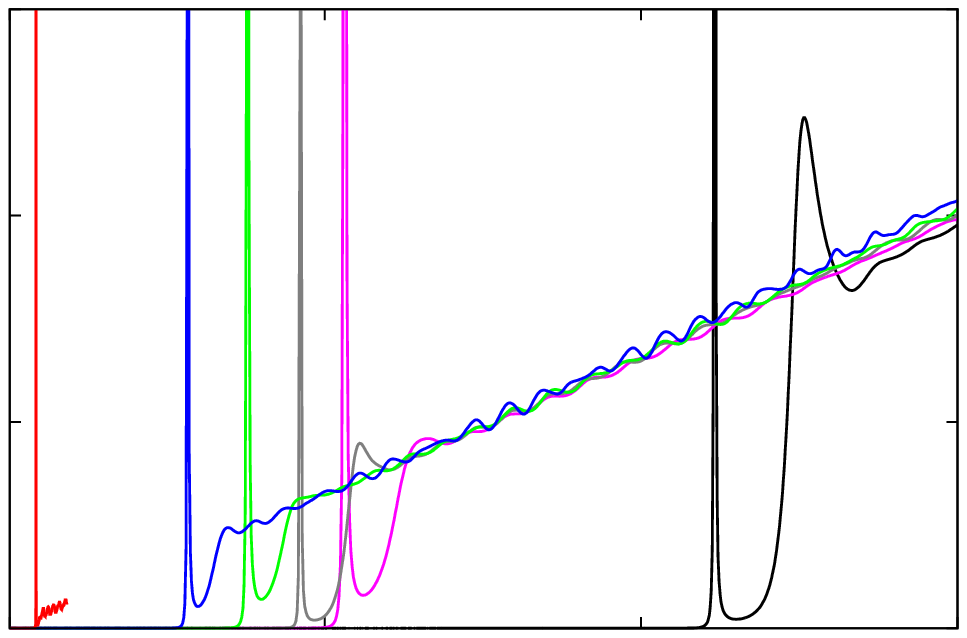}}%
    \gplfronttext
  \end{picture}%
\endgroup

   \caption{Plot showing $\text{Re}(\sigma)$ measured at $T=T_c/4$, $m^2=-4/L_e^2$ and $\kappa^2=0$ for a range of values of $\alpha$.  From left to right: red, $\alpha=-100$; blue, $\alpha=-1.0$; green, $\alpha=-0.25$; grey, $\alpha=0$; purple, $\alpha=0.125$ and black, $\alpha=0.24999$. The small oscillations are a numerical artefact.}
  \label{QuasinormalvaryAlpha}
}
This plot shows that increasing or decreasing $\alpha$ does not seem to hinder the development of these quasi-normal modes.  The dominant effect of, say, increasing the the GB coupling constant is to shift the poles to higher $\omega/T_c$.  This increase with $\alpha$ is particularly marked as you approach the upper limit of the coupling constant.

 \FIGURE{
   \centering
\begingroup
  \makeatletter
  \providecommand\color[2][]{%
    \GenericError{(gnuplot) \space\space\space\@spaces}{%
      Package color not loaded in conjunction with
      terminal option `colourtext'%
    }{See the gnuplot documentation for explanation.%
    }{Either use 'blacktext' in gnuplot or load the package
      color.sty in LaTeX.}%
    \renewcommand\color[2][]{}%
  }%
  \providecommand\includegraphics[2][]{%
    \GenericError{(gnuplot) \space\space\space\@spaces}{%
      Package graphicx or graphics not loaded%
    }{See the gnuplot documentation for explanation.%
    }{The gnuplot epslatex terminal needs graphicx.sty or graphics.sty.}%
    \renewcommand\includegraphics[2][]{}%
  }%
  \providecommand\rotatebox[2]{#2}%
  \@ifundefined{ifGPcolor}{%
    \newif\ifGPcolor
    \GPcolortrue
  }{}%
  \@ifundefined{ifGPblacktext}{%
    \newif\ifGPblacktext
    \GPblacktexttrue
  }{}%
  \let\gplgaddtomacro\g@addto@macro
  \gdef\gplbacktext{}%
  \gdef\gplfronttext{}%
  \makeatother
  \ifGPblacktext
    \def\colorrgb#1{}%
    \def\colorgray#1{}%
  \else
    \ifGPcolor
      \def\colorrgb#1{\color[rgb]{#1}}%
      \def\colorgray#1{\color[gray]{#1}}%
      \expandafter\def\csname LTw\endcsname{\color{white}}%
      \expandafter\def\csname LTb\endcsname{\color{black}}%
      \expandafter\def\csname LTa\endcsname{\color{black}}%
      \expandafter\def\csname LT0\endcsname{\color[rgb]{1,0,0}}%
      \expandafter\def\csname LT1\endcsname{\color[rgb]{0,1,0}}%
      \expandafter\def\csname LT2\endcsname{\color[rgb]{0,0,1}}%
      \expandafter\def\csname LT3\endcsname{\color[rgb]{1,0,1}}%
      \expandafter\def\csname LT4\endcsname{\color[rgb]{0,1,1}}%
      \expandafter\def\csname LT5\endcsname{\color[rgb]{1,1,0}}%
      \expandafter\def\csname LT6\endcsname{\color[rgb]{0,0,0}}%
      \expandafter\def\csname LT7\endcsname{\color[rgb]{1,0.3,0}}%
      \expandafter\def\csname LT8\endcsname{\color[rgb]{0.5,0.5,0.5}}%
    \else
      \def\colorrgb#1{\color{black}}%
      \def\colorgray#1{\color[gray]{#1}}%
      \expandafter\def\csname LTw\endcsname{\color{white}}%
      \expandafter\def\csname LTb\endcsname{\color{black}}%
      \expandafter\def\csname LTa\endcsname{\color{black}}%
      \expandafter\def\csname LT0\endcsname{\color{black}}%
      \expandafter\def\csname LT1\endcsname{\color{black}}%
      \expandafter\def\csname LT2\endcsname{\color{black}}%
      \expandafter\def\csname LT3\endcsname{\color{black}}%
      \expandafter\def\csname LT4\endcsname{\color{black}}%
      \expandafter\def\csname LT5\endcsname{\color{black}}%
      \expandafter\def\csname LT6\endcsname{\color{black}}%
      \expandafter\def\csname LT7\endcsname{\color{black}}%
      \expandafter\def\csname LT8\endcsname{\color{black}}%
    \fi
  \fi
  \setlength{\unitlength}{0.0500bp}%
  \begin{picture}(6802.00,4534.00)%
    \gplgaddtomacro\gplbacktext{%
      \csname LTb\endcsname%
      \put(814,704){\makebox(0,0)[r]{\strut{}0}}%
      \put(814,1892){\makebox(0,0)[r]{\strut{}50}}%
      \put(814,3081){\makebox(0,0)[r]{\strut{}100}}%
      \put(814,4269){\makebox(0,0)[r]{\strut{}150}}%
      \put(946,484){\makebox(0,0){\strut{}0}}%
      \put(2760,484){\makebox(0,0){\strut{}10}}%
      \put(4582,484){\makebox(0,0){\strut{}20}}%
      \put(6405,484){\makebox(0,0){\strut{}30}}%
      \put(176,2486){\rotatebox{-270}{\makebox(0,0){\strut{}$\sigma/T_c$}}}%
      \put(3675,154){\makebox(0,0){\strut{}$\omega/T_c$}}%
    }%
    \gplgaddtomacro\gplfronttext{%
    }%
    \gplbacktext
    \put(0,0){\includegraphics{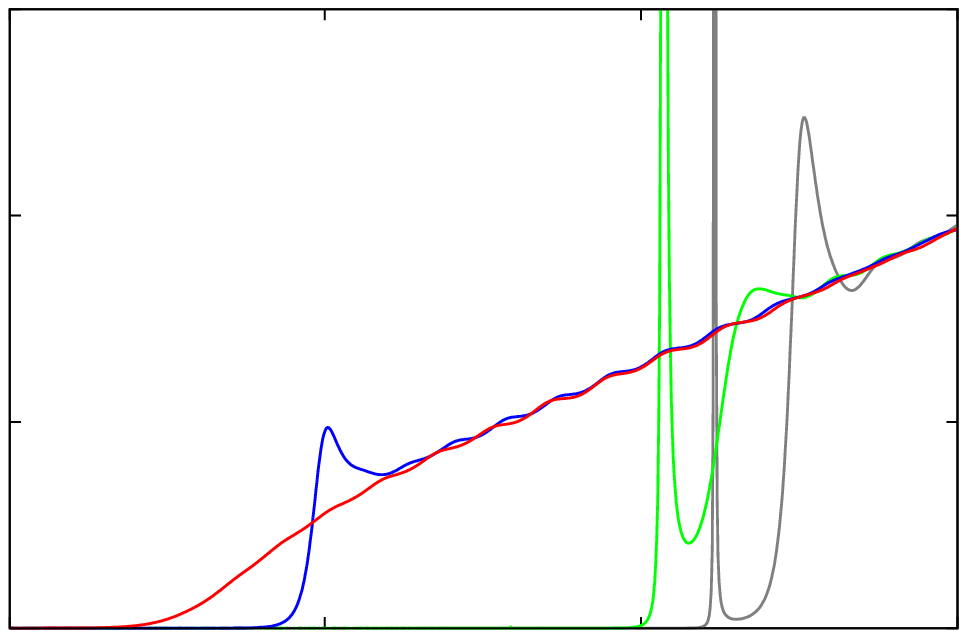}}%
    \gplfronttext
  \end{picture}%
\endgroup

   \caption{Plot showing $\text{Re}(\sigma)$ measured at $T=T_c/4$ at $m^2=-4/L_e^2$ and $\alpha=0.24999$ for a range of values of $\kappa^2$. From left to right: red $\kappa^2=0.1$; blue $\kappa^2=0.01$; green, $\kappa^2=0.001$ and grey, $\kappa^2=0.0001.$}
\label{QuasinormalvaryKappa}
}

Figure \ref{QuasinormalvaryKappa} shows the effect that backreaction has on the development of the quasi-normal modes with a plot of $\text{Re}(\sigma)$, measured at $T=T_c/4$, for $m^2=-4/L_e^2$, $\alpha=0.24999$ for a variety of $\kappa^2$.  We see that turning on backreaction very quickly removes the appearance of the poles, at least at this temperature; it is still quite conceivable that they may appear as the temperature is dropped.  Analysis of this phenomenon at much lower values of $\alpha$ show the existence of quasi-normal modes up to much higher values of $\kappa^2$, supporting the observation that the effect of backreaction diminishes as $\alpha$ is reduced.


\section{Conclusion}

The aim of this paper was to explore the dependence of the fully backreacting Gauss-Bonnet holographic superconductor on the mass of the scalar field.  We began by studying the critical temperature, $T_c$, of the system.  We found that in the majority of parameter space the effect of backreaction is to reduce $T_c$  but that in a narrow region where $m^2\to-4/L_e^2$ and $\alpha\to L^2/4$ its effect is reversed and actually increases $T_c$.  In this regime large, super-planckian values of  backreaction are numerically attainable.  We also found that as $\alpha$ becomes large and negative $T_c$ increases and the effect of backreaction is diminished as the gravitational action is dominated by the higher curvature terms.  Again, this provides a regime where large values of backreaction are attainable.  

We studied the zero temperature limit, proving that the system does not permit regular solutions with a tachyonic scalar field and place strict constraints on positive mass solutions.  Following \cite{Horowitz:2009ij} we relaxed the regularity constraint for a system without a black hole and permitted the fields to have logarithmic divergences.  Such a solution was found in the Einstein limit but was shown to be inconsistent with the idea of Gauss-Bonnet gravity as a perturbative expansion of Einstein gravity.  These findings show that a satisfactory zero temperature solution to this system has not yet been found and raises questions as to whether one can be found, except possibly if the scalar is massless.  

We also studied the conductivity of the system.   In the region away from the BF bound we saw that the prominent effect of, say, increasing $\alpha$ was to increase the frequency gap, $\omega_g$.  In the vicinity of the BF bound, the effect of $\alpha$ was to shift the location of the quasi-normal modes that appear there, again shifting them to larger $\omega$ as $\alpha$ was increased.  Otherwise $\alpha$ did not seem to effect their development.  The effect of backreaction was more notable: increasing backreaction away from the probe limit quickly prevented the appearance of these quasi-normal modes, at least within the temperature range that we were able to study.

The findings of this research suggest a number of interesting avenues for future research, most notably in relation to the zero temperature limit.   The first thing to do might be to either find, or disprove the existence of, positive mass solutions in this limit.  If they can be found it would be very interesting to study the quantum phase transitions that may happen there.  

Having shown that this system does not permit regular, superconducting, zero temperature, tachyonic solutions, it would be interesting to see if there exist non-regular solutions that are compatible with the perturbative relation between Einstein and GB gravity.  If so, it would be interesting to find out how this non-regularity can be interpreted.  It would also be interesting to study the conductivity of this system in such a case, particularly in light of  \cite{Siopsis:2010pi} which suggests that an infinite tower of quasi-normal modes may appear on the real axis as the temperature goes to zero.

Another interesting line of enquiry would be to further investigate the regime of positive scalar masses at finite temperature.  In this work we were only able to find solutions for very small $m^2$ which proved to be only marginally different to those of $m^2=0$.  In \cite{Kim:2009kb}, which studied the $m^2>0$ case in the probe limit, the authors found solutions at larger masses but found that solutions become difficult to attain, possibly due to an observed warping  of the solution space which was dramatically enhanced by increasing $m^2$.  It would be interesting to see how the inclusion of higher curvature terms and backreaction may affect this phenomenon.

Another interesting outcome of this paper is the observation that there exists two regions of parameter space where the numerical system is substantially more stable to the inclusion of backreaction.  Since the inclusion of backreaction is an important and often complicated aspect of these systems, these regimes may provide testing grounds where the numerical analysis of backreaction is somewhat easier.

\section*{Acknowledgements}
I would like to thank Ruth Gregory for very helpful input during this work.  I would also like to thank Sugumi Kanno and Paul Sutcliffe for previous collaboration and Danny Brattan, Simon Gentle and Laura Imrie for useful discussions.  This work is supported by an STFC studentship.

%
%
%
%
\providecommand{\href}[2]{#2}\begingroup\raggedright\endgroup

 \end{document}